\documentclass[twocolumn]{emulateapj} 
\usepackage{epsfig}

\newcommand{\msun}{\hbox{$\hbox{M}_{\odot}$}}

\newcommand{\HI}{\hbox{{\rm H}\kern 0.1em{\sc i}}}
\newcommand{\HII}{\hbox{{\rm H}\kern 0.1em{\sc ii}}}
\newcommand{\VI}{\hbox{$V_{606}$-$I_{814}$}}
\newcommand{\Vhst}{\hbox{$V_{606}$}}
\newcommand{\Ihst}{\hbox{$I_{814}$}}
\newcommand{\HIdens}{\hbox{$\langle$M$_{\mbox{\scriptsize{\HI}}}$/A$_{{\rm tail}}\rangle$}}
\newcommand{\mhi}{\hbox{M$_{\mbox{\scriptsize{\HI}}}^{{\rm tot}}$}}
\newcommand{\SCC}{\hbox{$\Sigma_{{\rm SCC}}$}}
\newcounter{figcount}
\shorttitle{Star Clusters in Tidal Tails}
\shortauthors{B.\ Mullan et al.\,}

\begin{document}


\title{Star Clusters in the Tidal Tails of Interacting Galaxies: Cluster Populations Across a Variety of Tail Environments} 

\author{B.\ Mullan\altaffilmark{1}, I.\ S.\ Konstantopoulos\altaffilmark{1}, A.\ A.\ Kepley\altaffilmark{2,}\altaffilmark{3}, K.\ H.\ Lee\altaffilmark{1}, J.\ C.\ Charlton\altaffilmark{1}, K.\ Knierman\altaffilmark{4}, N.\ Bastian\altaffilmark{5}, R.\ Chandar\altaffilmark{6} , P.\ R.\ Durrell\altaffilmark{7}, D.\ Elmegreen\altaffilmark{8}, J.\ English\altaffilmark{9}, S.\ C.\ Gallagher\altaffilmark{10}, C.\ Gronwall\altaffilmark{1}, J.\ E.\ Hibbard\altaffilmark{3},\,\, S.\ Hunsberger\altaffilmark{1}, K.\ E.\ Johnson\altaffilmark{2}, A.\ Maybhate\altaffilmark{11},\,\, C.\ Palma\altaffilmark{1}, G.\ Trancho\altaffilmark{12}, W.\ D.\ Vacca\altaffilmark{13}}

\altaffiltext{1}{Pennsylvania State University, Department of Astronomy \& Astrophysics, 525 Davey Lab University Park PA 16803; mullan@astro.psu.edu}
\altaffiltext{2}{Department of Astronomy, University of Virginia, University of Virginia, P.O. Box 400325, Charlottesville, VA 22904-4325}
\altaffiltext{3}{National Radio Astronomy Observatory, 520 Edgemont Road, Charlottesville, VA 22903-2475}
\altaffiltext{4}{Arizona State University, School of Earth and Space Exploration, Bateman Physical Sciences Center F-wing Room 686,Tempe, AZ 85287-1404}
\altaffiltext{5}{School of Physics, University of Exeter, Stocker Road, Exeter EX4 4QL}
\altaffiltext{6}{The University of Toledo, Department of Physics and Astronomy, 2801 West Bancroft Street, Toledo, OH 43606}
\altaffiltext{7}{Youngstown State University, Department of Physics and Astronomy, Youngstown, OH 44555}
\altaffiltext{8}{Vassar College, Department of Physics \& Astronomy, Box 745, Poughkeepsie, NY 12604}
\altaffiltext{9}{University of Manitoba, Department of Physics and Astronomy, Winnipeg, Manitoba R3T 2N2, Canada}
\altaffiltext{10}{The University of Western Ontario, Department of Physics and Astronomy, 1151 Richmond Street, London, Ontario, N6A 3K7, Canada }
\altaffiltext{11}{Space Telescope Science Institute, 3700 San Martin Drive, Baltimore, MD 21218}
\altaffiltext{12}{Gemini Observatory, Casilla 603, Colina el Pino S/N, La Serena, Chile}
\altaffiltext{13}{Stratospheric Observatory for Infrared Astronomy/ Universities Space Research Association, NASA Ames Research Center, MS 144-2, Moffett Field, CA 94035}

\slugcomment{Accepted for publication in the Astrophysical Journal}


\begin{abstract}

We have searched for compact stellar structures within 17 tidal tails in 13 different interacting galaxies using \textit{F606W}- and \textit{F814W}- band images from the Wide Field Planetary Camera 2 (WFPC2) on the \textit{Hubble Space Telescope} (\textit{HST}). The sample of tidal tails includes a diverse population of optical properties, merging galaxy mass ratios, \HI\ content, and ages. Combining our tail sample with Knierman et al.\ (2003), we find evidence of star clusters formed \textit{in situ} with M$_V <$ -8.5 and $V$-$I$ $<$ 2.0 in 10 of 23 tidal tails; we are able to identify cluster candidates to M$_V$ = -6.5 in the closest tails. Three tails offer clear examples of ``beads on a string" star formation morphology in $V$-$I$ color maps. Two tails present both tidal dwarf galaxy (TDG) candidates and cluster candidates. Statistical diagnostics indicate that clusters in tidal tails may be drawn from the same power-law luminosity functions (with logarithmic slopes $\approx$ -2 -- -2.5) found in quiescent spiral galaxies and the interiors of interacting systems. We find that the tail regions with the largest number of observable clusters are relatively young ($\lesssim$ 250 Myr old) and bright ($V \lesssim$ 24 mag arcsec$^{-2}$), probably attributed to the strong bursts of star formation in interacting systems soon after periapse. Otherwise, we find no statistical difference between cluster-rich and cluster-poor tails in terms of many observable characteristics, though this analysis suffers from complex, unresolved gas dynamics and projection effects.

\end{abstract}




\section{Introduction}

Galaxy interactions and mergers are commonly observed phenomena, from intermediate (z $\sim$ 1) to low redshifts. The complex gravitational potential of interacting galaxies drastically affects their morphologies, producing tidal tails and other disturbed features (\citealp{toomre}; \citealp{schweizer78}). Encounters between galaxies are also agents of photometric evolution, often localized a few kiloparsecs from the nucleus \citep{schombert} or strewn across tidally distorted features (\citealp{hibvan}; \citealp{duc}; \citealp{K07}; \citealp{schombert}; \citealp{hibbard05}). The intensity of these events may range from relatively rare global starbursts (e.g.\ NGC 6240 and Arp 220), to more frequent local concentrations of star-forming behavior that have little impact on the integrated optical colors of the host galaxies \citep{bergvall03}.  

Tidal tails are physically interesting environments for star formation. They are prevalent for $\sim$ 500 Myr--1 Gyr from their inception at the galaxies' initial encounter, before dispersing into the intergalactic medium \citep{binney08}. They can host alternating sequences of tidally compressive and extensive regions \citep{renaud09} that may shape any observed characteristics of emerging cluster populations, and are bereft of the familiar periodic density waves of spiral arms that can trigger new generations of clusters. Moreover, the low stellar and gaseous masses and densities of tidal tails (e.g. \citealp{elmegreen98}) compared to their progenitor galaxies ensures their star formation histories are easier to disentangle than the often heavily extincted, star-forming engines of interacting galaxy interiors. Tail internal extinctions are low (A$_V$ $\lesssim$ 0.5; e.g.\ \citealp{temporin05}), simplifying the photometric analysis often made arduous and convoluted by intricate extinction maps and inconvenient galaxy orientations.  

The resulting low stellar densities present challenges in studying the sparse stellar populations within tidal debris. One way to circumvent this issue is to focus on the clustered stellar component of tidal tails. With the observational resources of \textit{HST}, star clusters are luminous tracers of the tidal tail star formation history (\citealp{iraklis09}; \citealp{RdG09}) out to distances $\approx$ 10--70 Mpc. Thus they allude to the overall star formation capacity of their tidal homes and the interplay of physical properties that allow such clusters to form and survive.

Star clusters have been observed in tidal debris. For instance, populations of 40--50 compact clusters are found in the tidal tails of the Tadpole and Mice mergers (\citealp{RdG}; \citealp{tran}). \citet{bastian05} also find clusters in the tails of NGC 6872, which appear to have luminosity and mass functions similar to those of clusters within the main bodies of mergers. It is difficult to assess the prevalence of star clusters across a variety of tidal tails, however, because studies have different resolution and completeness limits, and there are strongly age-dependent effects of cluster evolution (i.e.\ fading). Furthermore, the connection between star formation in clusters and many observed features of galaxy mergers -- e.g.\ age, multiwavelength brightness and colors, and dynamical properties -- is not straightforward. For instance, \citet{K03}, hereafter K03, do not find a clear relationship between the number of star clusters they detect in tidal tails and various properties of a galaxy merger. They identify dozens of compact clusters brighter than M$_V$ = -8.5, with $V$-$I$ colors ranging from 0.2 to 0.9 in one system (NGC 3256), but do not find such an excess in three other galaxies (NGC 4038, NGC 3921, and NGC 7252). Their sample of interactions is limited to relatively old ($\approx$ 400 -- 730 Myr), optically bright major mergers, however. It is uncertain what this implies for merging galaxies in the full spectrum of interaction ages, mass ratios, and other observable characteristics.  

Tidal tails do not appear to consistently have star clusters, unlike in the disks of gas-rich mergers (\citealp{miller97}; \citealp{whitmore99}; \citealp{zepf99}). The single tidal tail with a confirmed cluster excess from K03 (NGC 3256W) has 0.11 $\pm$ 0.03  cluster candidates kpc$^{-2}$ for sources with $V$-$I$ $<$ 0.7 and M$_{V} <$ -8.5. Additionally, clusters found in tidal debris often have ages similar to young clusters found within the interacting galaxies. In the eastern tail of NGC 3256, K03 infer young ages from their broadband colors, and \citet{trancho} spectroscopically verify that several of these clusters are $\lesssim$ 200 Myr old. These ages are also younger than the 400 Myr age of the tail they inhabit, strongly suggesting \textit{in situ} formation. \citet{peterson09} identify clusters in the tidal bridge of the NGC 7714/15 system, whose ages are similar to clusters they find in the interacting galaxies. Unlike these interior clusters, however, they do not find many bridge objects with masses above their age-averaged mass completeness limit of 10$^{4.8}$ \msun.  

The dependence of clustered star formation on \HI\ properties is also not apparent. \citet{aparna07} find that a threshold of \HI\ column density N$_{{\mbox{\scriptsize{\HI}}}} \sim$ 10$^{20.6}$ cm$^{-2}$ is a \textit{necessary} but not \textit{sufficient} condition for  generating clusters at the M$_V <$ -8.5 level. It is true that projection effects may confuse the translation from column density to \HI\ volume density and pressure, but this questions how the complex interplay of \HI\ content, tail densities, pressures, etc.\ shapes the star-forming environments of tidal tails. 

K03 also find that some systems devoid of star clusters seem to prefer harboring larger star- or cluster-forming complexes whose luminosities and sizes are consistent with tidal dwarf galaxies (TDGs). TDGs are potentially self-gravitating dwarf galaxy-sized accumulations of tidal debris (\citealp{schweizer78}; \citealp{bh92}; \citealp{duc04}). NGC 7252 and NGC 3921 were presented in this context; they both show 7--8 clusters associated with their tidal dwarfs, and no statistical in-tail cluster excess anywhere else. The canonical example of a TDG is in NGC 4038/9 (\citealp{saviane04}; \citealp{hibbard01}), which shows evidence for star formation 2 Myr ago \citep{mirabel92}, or hundreds of Myr after the main tail formed. Through numerical simulations, \citet{hibbard95} and \citet{bournaud06} assert that many TDGs may live long enough to become independent from their host tails. Through H$\alpha$ Fabry-Perot observations and modelling, \citet{bournaud04} find that the TDG candidate of NGC 7252 is a real condensation of matter and not a chance projection effect.  \citet{boquien10} explore the SEDs and star-forming properties of this TDG (and others) in depth. 

It is possible that the apparent mutual exclusivity of compact clusters vs.\ TDGs identified in the K03 tails stems from small sample statistics, but it is also plausible that the outcome is affected by certain dynamical properties. For instance, the tail velocity dispersion and the resulting ambient pressure may conceivably encourage or discourage the formation of TDGs (\citealp{elm93}; \citealp{elm97}) or other bound structures. K03 posit that physical conditions in the tail may spur different modes of star formation, e.g.\ either TDG or tail cluster formation, but not both. Modeling work by \citet{duc04} suggests that variations in dark matter halo structure and other interaction parameters are key in deciding tail lengths and kinematic properties encouraging either star cluster formation along the tail or concentrated in kinematically distinct TDGs at the tail tip. This implies a top-down scenario where several gross tail and merger properties have the potential to help determine the cluster population in tidal debris.  Observationally, this hypothesis is limited to a small interaction sample (K03), and requires follow-up with additional interactions. 

Fundamentally, there may be environmental factors that dictate how effectively stars and star clusters form in interacting pair of galaxies. Evidence now suggests that the majority of star formation occurs in clusters (\citealp{lada}; \citealp{fall}; although see \citealp{bressert10}). While the formation efficiency of bound clusters appears to scale directly with the total star and cluster formation rate of its environment \citep{silvavilla10}, star clusters (observed and inferred) typically constitute $\lesssim$ 3--10\% of the stellar complement of evolved galaxies. Thus, most clusters must fall prey to a variety of disruption mechanisms and contribute to the dominant field star component. It is unknown how processes determining star cluster formation and survival, and the luminous subset of these clusters we observe, relate to the physical conditions within tidal tails and the large-scale observable properties (luminosity, color, mass, TDGs, etc.) they prescribe.

Cumulatively, these recent insights into the mechanics of clustered star formation and galaxy interactions merit a re-examination of clusters in tidal debris, specifically in terms of the characterization of these clusters and their relationships to their tail environments. In this paper, we extend the sample of merging galaxies of K03 with \textit{HST} WFPC2 \textit{F606W}- and \textit{F814W}- band observations of twelve additional galaxies. The seventeen tidal tails of these otherwise isolated systems were selected to test environmental extremes for star formation across a broad range of interactions.  An overview of the observations and data reduction is provided in \S 2, along with object detection and photometry. In \S 3, the properties of star cluster candidates of the tails are discussed, and \S 4 follows with a discussion of star cluster populations in the context of the observed tidal tail environment. \S 5 offers conclusions based on current data. Lastly, the appendices highlight observational details of the individual tidal tails and our efforts in optimizing the number of clusters detected at faint magnitudes across the sample.



\section{Observations and Reductions}

\subsection{Observations} 

Twelve interacting galaxies with seventeen tidal tails were selected for observation with the \textit{HST} WFPC2 in the Cycle 16 program 11134 (P.I.\ K.\ Knierman). This sample reflects an expansion of the K03 sample of six tidal tails (from four individual galaxies), as well as the parameter space of interaction characteristics -- interaction age, progenitor galaxy mass ratios, \HI\ and optical properties -- they define. The ``\HI\ Rogues Gallery" was used for selection guidance \citep{HIrogues}. The tidal tails selected in this study are displayed in \mbox{Figure 1}, as SDSS $r$-band images with \textit{HST} WFPC2 footprints overlaid. In systems where there are two obvious, separated tails or debris regions, they are indicated by an additional letter after their names, corresponding to their orientations (E for ``East," W for ``West," etc.). All other tails are indicated by their primary catalog designations.

\begin{figure*}[htbp]
\plotone{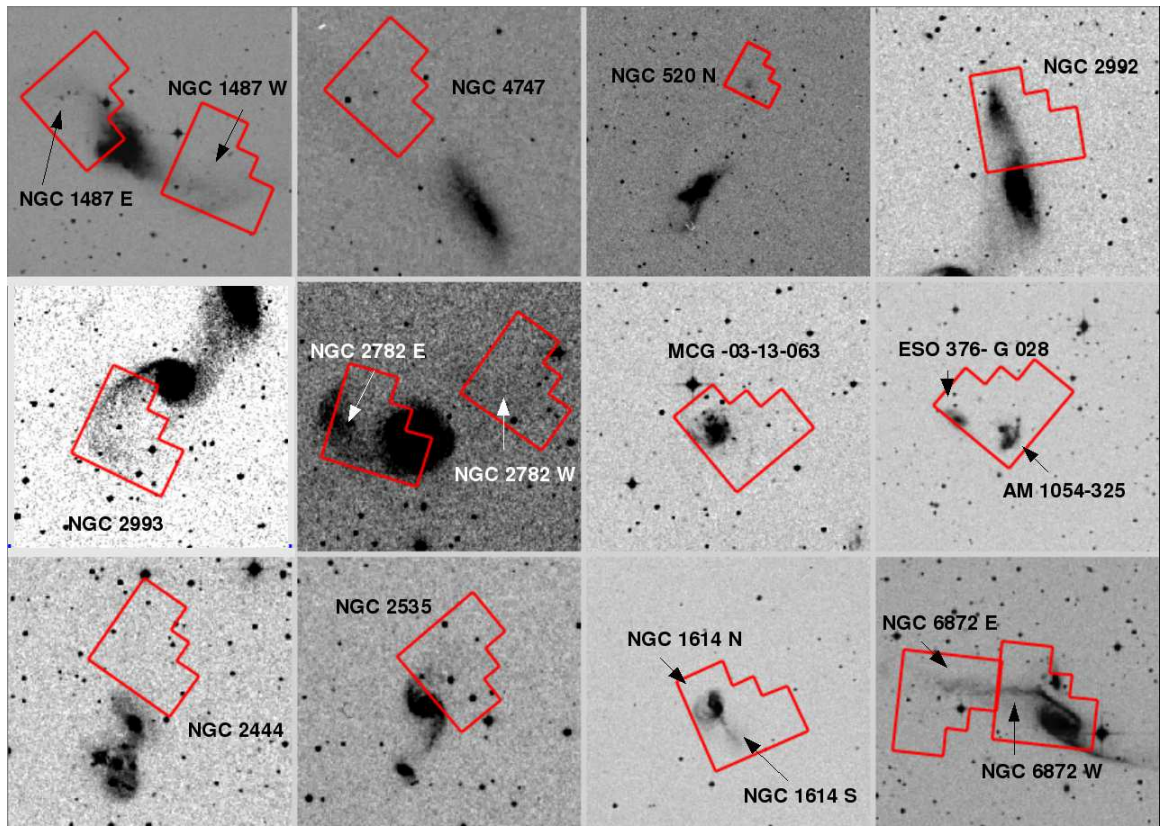}
\caption{Sloan Digital Sky Survey $r$-band images of the galaxies used in this project, with \textit{HST} fields of view overlaid. The distinctions between eastern and western or northern and southern tails are indicated by separate pointings (NGC 1487 and NGC 2782), or separate regions within a single image (NGC 1614). North and east are up and to the left, respectively. Images are stretched and scaled individually to enhance tidal features.}
\label{alltails}

\end{figure*}
 \addtocounter{figcount}{1}

 WFPC2 images were taken from 2007 September to 2008 August in both \textit{F606W} and \textit{F814W} (``$V_{606}$" and ``$I_{814}$" hereafter\footnotemark[1]); a log of these observations is given in \mbox{Table 1}. Exposure times, t$_{\rm{exp}}$, range from 900 to 2100 s in \Vhst\ and \Ihst, representing the total integration time of at least two separate exposures in each filter. This was meant to facilitate the identification and removal of cosmic rays. Images at two different dither positions were obtained for AM 1054-325/ESO 376- G 028, NGC 1614N/S, NGC 2444, NGC 2535, and NGC 6872E/W, with an offset of 2.5 and 5.5 pixels in the WF and PC chips, respectively. Integration times were selected to equal image depths of K03, allowing for direct comparison of analyses and results. In all cases, a gain of 7 e$^{-1}$ ADU$^{-1}$ was used.  

\footnotetext[1]{$V$ and $I$ will refer to corresponding Johnson-Cousins magnitudes.}



\begin{table}[htbp]
\begin{center}
{\scriptsize
\begin{tabular}{llrl} 
\multicolumn{4}{c}{\sc Table 1. Journal of Observations} \\
\hline\hline 
Merger/Tail & Filter & t$_{\rm{exp}}$ (s) & Date \\
\hline 
NGC 1487E & F606W & 1000 & 2008 Aug 9 \\
 	 & F814W & 900 & 2008 Aug 9 \\
NGC 1487W & F606W & 1000 & 2008 Aug 31 \\
 	 & F814W & 900 & 2008 Aug 31 \\
NGC 4747 & F606W & 1000 & 2008 Jan 1 \\
 	 & F814W & 900 & 2008 Jan 1 \\
NGC 520 & F606W & 1000 & 2007 Sept 23 \\
 	 & F814W & 900 & 2007 Sept 23 \\
NGC 2992 & F606W & 1000 & 2007 Dec 28 \\
 	 & F814W & 900 & 2007 Dec 28 \\
NGC 2993 & F606W & 1000 & 2007 Dec 3 \\
 	 & F814W & 900 & 2007 Dec 3 \\
NGC 2782E & F606W & 1000 & 2007 Oct 8 \\
 	 & F814W & 900 & 2007 Oct 8 \\
NGC 2782W & F606W & 1000 & 2007 Nov 8 \\
 	 & F814W & 900 & 2007 Nov 8 \\
MCG-03-13-063 & F606W & 1000 & 2007 Nov 24 \\
 	 & F814W & 900 & 2007 Nov 24 \\
AM 1054-325 & F606W & 1900 & 2008 Feb 24 \\
 	 & F814W & 1900 & 2008 Feb 24 \\
NGC 2444 & F606W & 1900 & 2007 Oct 13 \\
 	 & F814W & 1900 & 2007 Oct 13 \\
NGC 2535 & F606W & 1900 & 2007 Dec 9 \\
 	 & F814W & 1900 & 2007 Dec 9 \\
NGC 1614N/S & F606W & 1900 & 2007 Nov 15 \\
 	 & F814W & 1900 & 2007 Nov 15 \\
NGC 6872E & F606W & 2100 & 2008 Feb 23 \\
 	 & F814W & 2100 & 2008 Feb 23 \\
NGC 6872W & F606W & 1900 & 2008 May 16 \\
 	 & F814W & 1900 & 2008 May 16 \\

\hline

\end{tabular}}
\end{center}
\end{table}



Pairs of images were reduced, averaged with the IRAF\footnotemark[2] task GCOMBINE, and cleaned for hot pixels with COSMICRAYS. Moreover, the low bias problem of WF4 has only been corrected for NGC 1487E/W. According to \citet{aparna08}, this problem is manifested in image streaking whose photometric corrections are most pronounced in faint objects. Given this work's preferential avoidance of this chip for tail placement, this issue would ordinarily only affect the tails of NGC 4747, NGC 2782, NGC 2535, and NGC 6872. However, our images for these systems employed deep exposures and are clean of image streaking (WF4 mean biases are on par with those of the WF2 and WF3 chips; $\gtrsim$ 300 DN). Thus, this will not have an appreciable effect on the data analysis. In addition, the Planetary Camera chip (PC) was not analyzed because source detection was compromised by a greater readnoise contribution from the larger number of pixels across each source.

\footnotetext[2]{IRAF is distributed by the National Optical Astronomy Observatory, operated by AURA, Inc., under contract to the National Science Foundation.}

Neutral hydrogen data for NGC 2782, NGC 2992/3, NGC 2444, and NGC 2535 were taken from the VLA archives\footnotemark[3]. The data were reduced in AIPS \citep{greisen10}, using standard procedures. The neutral hydrogen data for NGC 6872 and NGC 1487 were obtained from the ATCA archives\footnotemark[4] and calibrated using the standard procedures of the ATNF version of \textit{Miriad} \citep{miriad}. An upcoming publication that focuses on the \HI\ distribution of these tails on kpc scales (Mullan et al.\ 2011; in preparation) will document the reduction procedures in greater detail. \HI\ data for NGC 520 and NGC 1614 were obtained from \citet{hibvan} and \citet{hibbard96}, respectively. 

\footnotetext[3]{http://archive.nrao.edu/}
\footnotetext[4]{http://atoa.atnf.csiro.au/}

\subsection{Tail Definitions and Properties} 

Following K03, contiguous regions bounded by SAO DS9 contours of one count above the average background in $V_{606}$-band images were defined as ``in tail." We applied this criterion to histogram-equalized images smoothed with a Gaussian kernel with a 5--7 pixel (0.5--0.7\arcsec) FWHM. The tail surface brightness limit ranges between 24.7 -- 25.8 mag arcsec$^{-2}$ in $V$. All other regions were designated ``out of tail." Where galaxy interiors are imaged (e.g.\ AM 1054-325 E/W, MCG -03-13-063, NGC 2782E, and NGC 6872W), the inner extent of the tail was defined by DS9 contours set to where radial light profiles (measured by IRAF PVEC) showed a change in scale length as the disk connects to the tail. This method isolates ``anti-truncated" tails that may be the result of gravitational interactions \citep{erwin}.

For AM 1054-325, radial light profiles from the inner galactic region were azimuthally-averaged into an ellipse (center at 10:46:58.877, -33:08:23.60; position angle 64.95$^{\circ}$, major axis 5.989\arcsec, minor axis 5.661\arcsec). This system's tail may extend into this interior, but this superposition of tidal debris and host galaxy is unfit for our purposes. NGC 2444 is one final exception -- it has no optical tail evident in WFPC2 images. Consequently, its in-tail definition relies on its \HI\ debris. To be clear, all other tail definitions rely on said \textit{optically delineated} tails in order to allow direct comparison to K03. 

Tail regions were converted from SAO DS9 contours to IMAGEJ \citep{IJ} polygons for optical photometry. Regions were masked to exclude obvious foreground stars and resolved background galaxies. Mean count measurements and errors from IMAGEJ were converted to WFPC2 and Johnson-Cousins magnitudes using the zeropoints and filter transformations of \citet{cteweb} and \citet{holtzman95}. Corrections for Galactic extinction were made using A$_V$ values from \citet{schlegel}, transformed to the appropriate bandpasses according to the \citet{girardi} prescription. For this extinction treatment, we used an effective cluster temperature T$_{\rm{eff}}$ = 15,000 K, representing an aggregation of late B -- early A stars, of ages 10 -- 300 Myr for the input spectrum.

 \mbox{Table 2} presents a phenomenological summary of the tails. Tails are ordered by increasing luminosity distance (these values are provided in \S 3--4); we separate our sample from that of K03 in order to isolate any systematic errors that may accrue from the different bandpasses used in the observations. Many of these data from that paper are reproduced here for convenience. 
 
\begin{table*}[htbp]
{\scriptsize
 
\begin{tabular*}{1.0\textwidth}{@{\extracolsep{\fill}}lrrrrrrl|rrr}
\multicolumn{11}{c}{\sc Table 2. Properties of Tidal Tail Sample} \\
\hline\hline
\multicolumn{8}{l}{Global Properties} & \multicolumn{3}{l}{WFPC2 FOV Properties} \\
\hline
Tail & Interaction & Mass & Galaxy & SDSS Tail & \HI\ Tail  & M$_{\mbox{\tiny{\HI}}}^{{\rm tot~d}}$ & TDG & Tail $V^c$ &  Tail & $\langle$M$_{\mbox{\tiny{\HI}}}$/A$_{{\rm tail}}\rangle^{d}$ \\
 	& Age$^a$ & Ratio$^a$& SFR$^b$ &  Length$^c$ & Length$^{c,d}$ & (10$^9$ \msun) & Candidates$^e$ & (mag & $V-I^c$& (10$^6$ \msun\ \\
	& (Myr)  & & (\msun\ yr$^{-1}$)  &  (kpc)  &  (kpc) &  & & arcsec$^{-2}$) & & kpc$^{-2}$)  \\    
\hline
NGC 1487E & 500 & 0.25 & 0.12 $\pm$ 0.01 &        9 &       19 & 0.614$\pm$ 0.12 & no & 24.02 & 0.84 & 25.5$\pm$ 5.1 \\
NGC 1487W & 500 & 0.25 & 0.12 $\pm$ 0.01 &       12 &       17 & 0.541$\pm$ 0.11 & no & 24.57 & 1.10 & 33.2$\pm$ 6.6 \\
NGC 4747 & 320$^f$ & 0.05 & 0.27 $\pm$ 0.02 &       22 & - & - & no & 24.64 & 0.94 & - \\
NGC 520 & 300 & 0.05 & 6.5 $\pm$ 0.5 &      112 &      227 & 1.12$\pm$ 0.22 & yes & 24.21 & 1.04 & 3.81$\pm$ 0.76 \\
NGC 2992 & 100 &        1 & 3.3 $\pm$ 0.2 &       23 &       33 & 1.43$\pm$ 0.28 & yes & 23.47 & 1.14 & 6.44$\pm$ 1.3 \\
NGC 2993 & 100 &        1 & 5.0 $\pm$ 0.3 &       37 &       55 & 0.727$\pm$ 0.15 & no & 24.78 & 0.73 & 5.77$\pm$ 1.2 \\
NGC 2782E & 200 & 0.25 & 3.7 $\pm$ 0.3 &       37 &       17 & 2.41$\pm$ 0.48 & no & 23.81 & 1.02 & 7.08$\pm$ 1.4 \\
NGC 2782W & 200 & 0.25 & 3.7 $\pm$ 0.3 &       76 &       58 & 3.14$\pm$ 0.63 & no & 25.49 & 1.05 & 5.17$\pm$ 1.0 \\
MCG -03-13-063 & 100$^f$ & 0.25 & 0.43 $\pm$ 0.03 &       41 & - & - & no & 23.91 & 1.01 & - \\
ESO 376- G 028 & 85$^f$ &        1 & - &       10 & - & - & no & 22.76 & 1.33 & - \\
AM 1054-325 & 85$^f$ &        1 & 0.64 $\pm$ 0.04 &       16 & - & - & yes & 22.65 & 0.84 & - \\
NGC 2444 & 100 & 0.50 & - &      175 &      233 & 5.20$\pm$ 1.0 & no & - & - & 1.40$\pm$ 0.28 \\
NGC 2535 & 100 & 0.30 & 2.9 $\pm$ 0.2 &       25 &       79 & 4.45$\pm$ 0.89 & no & 24.37 & 1.02 & 6.84$\pm$ 1.4 \\
NGC 6872E & 150 & 0.20 & 2.9 $\pm$ 0.2 &       38 &       21 & 3.52$\pm$ 0.50 & no & 24.06 & 0.75 & 2.94$\pm$ 0.59 \\
NGC 6872W & 150 & 0.20 & 2.9 $\pm$ 0.2 &       21 &       21 & 3.52$\pm$ 0.50 & no & 23.57 & 1.16 & 1.68$\pm$ 0.34 \\
NGC 1614N & 750 &        1 & 35. $\pm$ 3. &       21 &       21 & 0.768$\pm$ 0.16 & no & 22.27 & 0.88 & 2.56$\pm$ 0.52 \\
NGC 1614S & 750 &        1 & 35. $\pm$ 3. &       21 &       21 & 0.964$\pm$ 0.19 & no & 23.00 & 0.68 & 2.89$\pm$ 0.59 \\
\hline
\multicolumn{11}{c}{The K03 Sample} \\
\hline
NGC 4038 & 420$^f$ &        1 & 5.2 $\pm$ 0.4 &       78 &       85 & 0.200$\pm$ 0.040 & yes & 24.54 & 1.05 & 1.09$\pm$ 0.22 \\
NGC 3256E & 400$^f$ &        1 & 33. $\pm$ 2. &       51 &       77 & 1.40$\pm$ 0.28 & no & 24.04 & 0.76 & 1.51$\pm$ 0.30 \\
NGC 3256W & 400$^f$ &        1 & 33. $\pm$ 2. &       46 &       57 & 2.20$\pm$ 0.44 & no & 23.75 & 0.70 & 2.96$\pm$ 0.59 \\
NGC 7252E & 730$^f$ &        1 & 5.4 $\pm$ 0.4 &       40 &       66 & 1.50$\pm$ 0.30 & yes & 25.42 & 0.89 & 0.593$\pm$ 0.12 \\
NGC 7252W & 730$^f$ &        1 & 5.4 $\pm$ 0.4 &       94 &      156 & 2.20$\pm$ 0.44 & yes & 24.46 & 1.09 & 0.573$\pm$ 0.11 \\
NGC 3921S & 460$^f$ &        1 & 1.9 $\pm$ 0.1 &       43 &       53 & 4.30$\pm$ 0.86 & yes & 24.72 & 0.94 & 0.872$\pm$ 0.17 \\

\hline\hline

\end{tabular*} 

{\footnotesize
$^{a}$When available, ages and mass ratios are adopted from the literature. We adopt errors in ages of $\pm$ 50 Myr and 25\% for mass ratio. See Appendix A for references. The mass ratio is defined as the mass of the perturbing galaxy divided by that of the main galaxy. 
\\
$^b$Galaxy SFRs based on \textit{IRAS} IR fluxes \citep{helou88}. See K03 and \citet{kennicutt98} for the details of the calculation. \textit{IRAS} measurements are unavailable for ESO 376- G 028 and NGC 2444.
\\
$^c$Tail photometry and length measurements were performed with IMAGEJ and corrected to standard $V$ and $I$ filters following \citet{holtzman95}; see \S 2.3. No optical tail detection or photometry is possible for NGC 2444, so no values are given. Errors are $\approx$ 0.1 for $V$ and 0.14 for $V$-$I$. 
\\
$^d$\HI\ measurements were performed as per \S2.2. Dashes indicate where \HI\ data are unavailable. \HI\ data and maps for K03 tails were obtained from \citet{aparna07} and references therein. 
\\
$^e$Based on visual inspection of SDSS and WFPC2 images. 
\\
$^f$These tails lacked age estimates in the literature, so these values were obtained by dividing the projected length of the tail by the estimated escape velocity. Similarly-calculated ages of K03 tails are reproduced from that publication. 
\\
}
}

\renewcommand{\thefootnote}{\arabic{footnote}}

\end{table*}

 
The first eight parameters listed pertain to ``global properties" of the interacting systems. Interaction ages (measured from periapse) and mass ratios are supplied by the literature; see Appendix A for specific references for each system. In cases where reliable ages were not found, they were estimated by dividing the projected length of the tail by an estimate of the escape or rotation velocity. This was demonstrated to provide acceptable accuracy ($\pm$ 50 Myr) in K03. Mass ratios are defined as the mass of the perturbing galaxy divided by the mass of the primary system. Table 2 also lists host galaxy SFRs based on \textit{IRAS} IR fluxes \citep{helou88}. K03 and \citet{kennicutt98} have details of the calculation.

Tail lengths were measured with freehand lines in IMAGEJ, tracing the tail curvature evident in SDSS $r$-band images and our \HI\ maps. Note that in many cases the \HI\ debris evidently extends beyond the optical tails (NGC 2782 is an interesting counterexample). Furthermore, we calculated the total \HI\ mass of entire tidal tails, \mhi. In every channel of the \HI\ data cubes, we measured the total \HI\ flux (Jy km s$^{-1}$) and summed the flux in each channel to get the total \HI\ flux F$_{\HI}^{\rm{tot}}$. The total \HI\ mass \mhi\ is then (e.g.\ \citealp{HIeqn}):
\begin{equation}
\mhi~[\msun] = 2.36 \times 10^5 ~(D[\mathrm{Mpc}])^2~\rm{F}_{\HI}^{\rm{tot}}[\mathrm{Jy~km~s}^{-1}]~,
\end{equation}
\noindent where $D$ is the distance to the tail in Mpc. We additionally indicate in \mbox{Table 2} which of these tails have visually identified TDG candidates.

The last three columns of \mbox{Table 2} list WFPC2 FOV-scale properties of the tidal tails. We include the WFPC2-imaged optical tail $V$-band surface brightness and $V-I$ color; for the K03 sample, tail photometry was conducted identically to our sample and corrected from the F555W/F814W bands to Johson-Cousins magnitudes (\citealp{holtzman95}; \citealp{cteweb}). Lastly, \mbox{Table 2} lists the average observed \HI\ surface density \HIdens\ within the optical tail areas (A$_{\rm{tail}}$) of the WFPC2 FOVs. \HIdens\ is expressed as:
\begin{eqnarray}
\nonumber \HIdens~[\msun~\mathrm{kpc}^{-2}] = 2.36 \times 10^5 ~(D[\mathrm{Mpc}])^2 \times \\
\rm{F}_{\HI}^{\rm{tail}}[\mathrm{Jy~km~s}^{-1}]/\rm{A}_{\rm{tail}}[\mathrm{kpc}^{2}]~.~
\end{eqnarray}
\noindent Here, F$_{\HI}^{\rm{tail}}$ is the \HI\ flux for the optical tail region within the WFPC2 FOV, and A$_{\rm{tail}}$ is the area of the optical tail region measured from the \HI\ images. This is similar to the area measured from the HST images, modulo differences in pixel sizes. The uncertainty in this quantity represents the uncertainties in measuring the flux of HI and assumes a 20\% error in measuring the distance. 

Thus, we have a ``global" gas-richness parameter \mhi\ and a WFPC2-scale \HI\ density metric \HIdens\ directly comparable to the optical counterparts and the populations of star clusters they may contain. Given the paucity of numerical modeling for many of these interactions, inclinations remain highly uncertain and therefore no attempts at adjusting projection-dependent quantities (e.g.\ \HIdens, tail lengths, and optical properties) have been made.

\subsection{Object Detection} 

Sources were detected using the DAOFIND algorithm in the IRAF package DAOPHOT. The criteria for detection were chosen to match K03 and provide a source list for photometry and analysis. These include the following:

\begin{itemize}
	\item At least 2 counts per object in \Vhst\ and \Ihst, along with a count threshold 2$\sigma$ above the local background. A sky annulus with 5--8 pixel radii was used for both measuring noise around each object and determining its sky background. 
	\item A signal-to-noise ratio S/N $>$ 3.
	\item An error in $V_{606}$-band magnitude $<$ 0.25.
	\item Detections in at least one dither position, accounting for the pixel offsets.
\end{itemize}

\noindent For every in- and out-of-tail region, obvious background galaxies or foreground stars were removed manually. 

\subsection{Photometry and Source Selection} 

Aperture photometry was performed on the source lists and cleaned images using the PHOT task in the package APPHOT. In light of potential source crowding, we used apertures with radii of 2, 5, and 8 for object, inner background, and outer background annuli, respectively.  Most importantly, these apertures were selected for consistency with K03. 

Such a small object aperture is not problematic. Even the closest systems in our sample would not likely have individual clusters marginally resolved by WFPC2; NGC 1487E/W and NGC 4747 images have a pixel scale of $\approx$ 5--10 pc, compared to probable r$_{eff}$ values $\sim$ 4 pc (\citealp{larsen04} and references therein) to $\sim$ 10 pc \citep{trancho}. Thus, before diffraction, an object aperture with a 2-pixel radius will encompass $>$ 1--2 effective radii of a cluster, or 50--100\% of its light, and is therefore acceptable.

Corrections for Galactic extinction were made with the procedure documented in \S 2.2. Nonoptimal charge transfer efficiency was corrected following \citet{cte}, with the formulation of \citet{cteweb}. Uncertainties for all values as well as photometric errors were propagated into the final values of all magnitudes and colors. Zeropoints were obtained from \citet{cteweb}.

Moreover, the concentration index $\Delta_V$ -- the difference in $V_{606}$ magnitude between apertures of 0.5 pixel and 3 pixel radii -- was calculated, with annuli of 6--9 pixel radii for background subtraction. Similar work can be found in \citet{whitmore93}, \citet{miller97}, and K03. The latter paper employs cutoff values for this structural parameter of 2.4 -- 3, depending on distance, to eliminate extended objects from their source lists and concentrate on point sources. Because clusters would be largely unresolved and there would be no physical link between $\Delta_V$ and the size of single clusters, no restrictions on $\Delta_V$ have been made here. Furthermore, we do not wish to exclude marginally resolved cluster complexes or young clusters embedded in nebular emission (H$\alpha$ is present in the \Vhst\ bandpass); these would appear more diffuse in $\Delta_V$ than individual compact clusters. We retain $\Delta_V$ data in Appendix A for reference.    

Aperture corrections of 0.33 in $V_{606}$ and 0.36 in $I_{814}$, based on the brightest 5--10 point sources in each chip on all images, were applied (this includes the 0.1 mag correction between 0.5$\arcsec$ and infinite apertures as per \citealp{holtzman95}). We calculate uncertainties of 0.008 and 0.01 for these values, respectively. Errors in using stellar aperture corrections vs.\ cluster candidate aperture corrections for the most cluster-deficient tails -- a few hundredths of a magnitude -- are negligible for this investigation.

Objects detected and measured with the above methodology were defined as star cluster candidates (SCCs) if they met the following color-magnitude criteria:
\begin{itemize}
 	\item $V$-$I$ $<$ 2.0. When translated into WFPC2 bandpasses, this color limit \VI\ $\approx$ 1.43. This criterion allows for populations of old, metal poor globular clusters ($V$-$I$ $\sim$ 1) akin to those observed in the Galaxy \citep{Reed}, as well as metal-rich clusters with $V$-$I$ $<$ 1.5 (\citealp{Peterson}; \citealp{Ajhar}, \citealp{whitmore95}; \citealp{Kundu}). This also includes globular clusters with $V$-$I$ $\sim$ 1.2--1.3 \citep{Lee} and solar metallicity. This color limit also allows clusters with redder colors from the stochastic populating of the stellar IMF at these low cluster masses \citep{maiz09}, additional extinction (even if overall A$_V \lesssim$ 0.5 mag), and uncertainties in metallicities. This $V$-$I$ limit is also needed to account for problems in modeling binarity in stellar populations, mass segregation, and other dynamical effects than can influence a cluster's integrated magnitudes and colors (see \citealp{pzwart10} for a review). Furthermore, \citet{anders} suggests that real clusters, subject to a variety of age- and environment-dependent disruption mechanisms, will fade and redden more rapidly with time than these traditional photometric models indicate, necessitating a more lenient color cutoff. 
	\item M$_V$ $<$ -8.5 ($\approx$ -8.6 in the WFPC2 system). To compare the SCC quantity across all tidal debris regions, a constant magnitude limit relatively unaffected by completeness is required. Most studies typically use such a bright magnitude cutoff (M$_V$ -8 -- -9: K03; \citealp{K07}, \citealp{schweizer96} \citealp{whitmore99}). This limit protects against a level of contamination from single main sequence and post main sequence stars prohibitive to statistical in-tail SCC identification \citep{efremov87}; Appendix B has more on this subject.
\end{itemize}

\noindent We find median errors in \Vhst\ and \Ihst\ of 0.08 and 0.15, respectively, for all sources. This includes contributions from photometric errors, CTE, and aperture correction uncertainties. Finally, despite identical absolute magnitude limits, K03 used a different $V$-$I$ $<$ 0.7 threshold to select SCCs. This color distinction was arbitrary, based on a comparison of the in-tail SCC colors with photometry reported for clusters within their host galaxies. The followup work of \citet{aparna07} employ an equally valid $V$-$I$ $<$ 2.0 criterion that also avoids selecting individual, luminous stars. In order to compare the results of those two benchmark works and combine with our own analyses, we chose the latter color and opt to recalculate many quantities published in K03 from their original source lists and their tails. We present them alongside our own in \S 3. 

\subsection{Completeness} 

Completeness curves are provided in \mbox{Figure 2} for optically-defined in-tail and out-of-tail regions, in both \Vhst\ and \Ihst. The completeness fraction for each tail was determined using 10,000 artificial stars, sampled 100 at a time with the ADDSTAR procedure in the package DAOPHOT. K03 and \citet{whitmore99} find similar completeness fractions for their samples derived from WFPC2 data; $\sim$ 50\% is reached at $V$ $\approx$ 25.7 (\Vhst\ $\approx$ 25.5). This corresponds roughly to a visual estimate for completeness from the color-magnitude diagrams of this project's sample (\S 3 and Figure Set 3). Completeness limits for out-of-tail regions are typically lower by 0.1--0.5 mag, i.e.\ by typical A$_V$ values for tidal tails. Beyond the tidal debris, detections are 50\% complete at $V$ $\approx$ 25.7--26.0, or \Vhst\ $\approx$ 25.6--25.9.

Translated to absolute magnitudes, these tail regions are 50\% complete to M$_V$ = -4.6 for the closest debris (NGC 1487E/W), to -8.6 for the most distant tails (NGC 1614N/S). The optically bright merger AM 1054-325, has the brightest absolute 50\% completeness limit of M$_V$ =-8.7. In \S 3.2, we discuss the ramifications of using a magnitude limit encumbered by the incompleteness of these tails.


\begin{figure*}[htbp]
\plotone{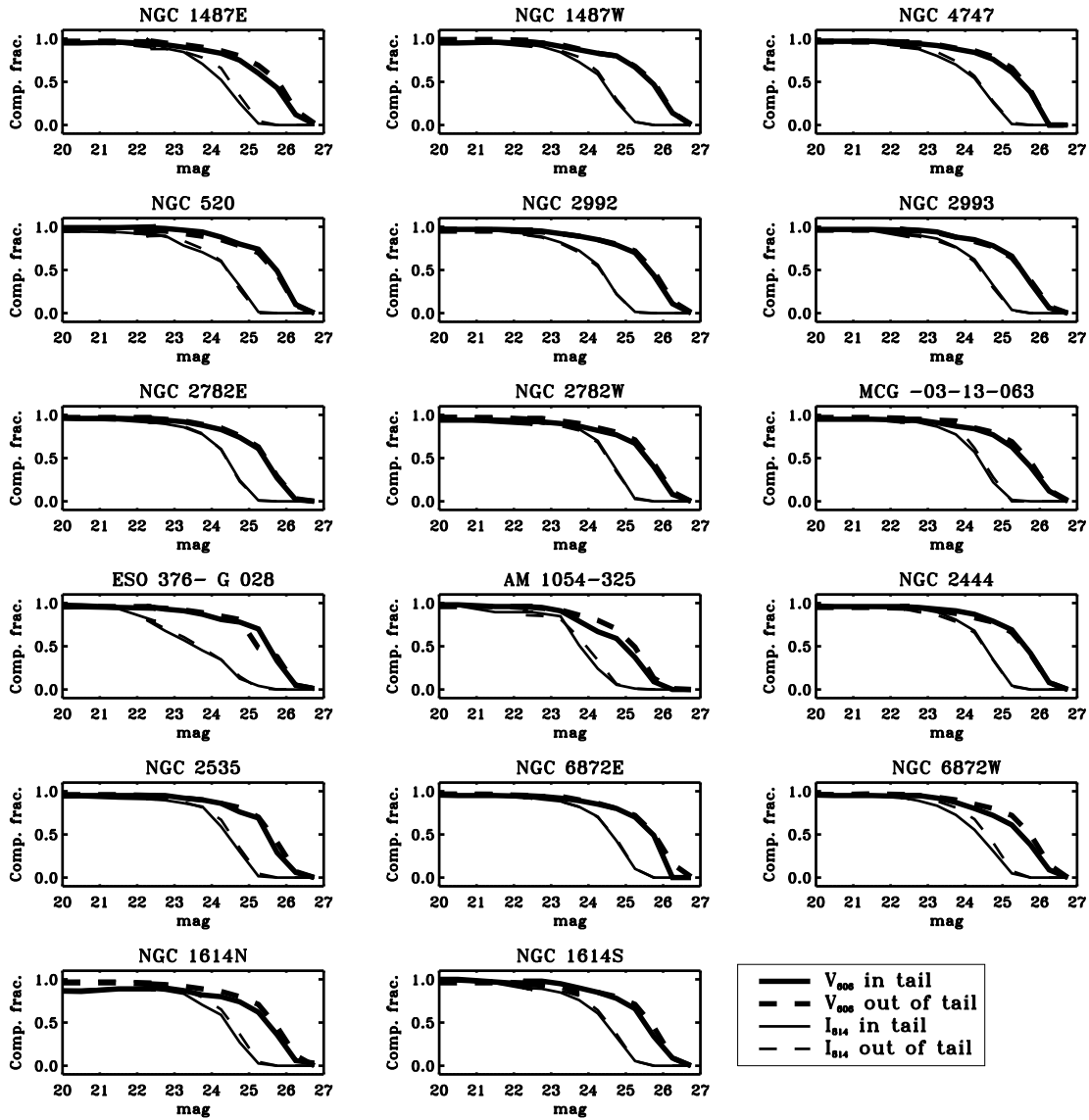}
\caption{\Vhst\ (thick lines) and \Ihst\ (thin lines) completeness curves computed for all WFPC2 pointings. Solid lines indicate calculations for in-tail regions, and dashed lines show the analysis of out-of-tail regions.}
\addtocounter{figcount}{1}
\end{figure*}




\section{Summary of the Tail Sample and Results}

In Appendix A, we summarize the results from this investigation for each tail of the sample and compare them to studies in the literature. As an example of this work, \mbox{Figure 3} introduces \HI\ contours on an SDSS $r$-band image (panel a), and a \Vhst-band mosaic of NGC 1487E\footnotemark[5], with in-tail and out-of-tail sources indicated (b). A background-subtracted \VI\ map of the tail region is also provided (c). This map was smoothed with a 12 pixel (1.2$\arcsec$) Gaussian kernel, corresponding to a linear size of $\approx$ 70--380 pc, depending on distance. This was done to circumvent the typically noisy, low signal registered by the in-tail pixels. Color-magnitude diagrams (CMDs) for in-tail and out-of-tail sources are displayed in this Figure (d -- e), as well as plots of \VI\ color vs. $\Delta_{V}$ (f). 

\footnotetext[5]{Similar Figures (3.1--3.17) for all tails are available in the electronic edition of {\it The Astrophysical Journal.}}


\begin{figure*}
\includegraphics[width=1.0\textwidth]{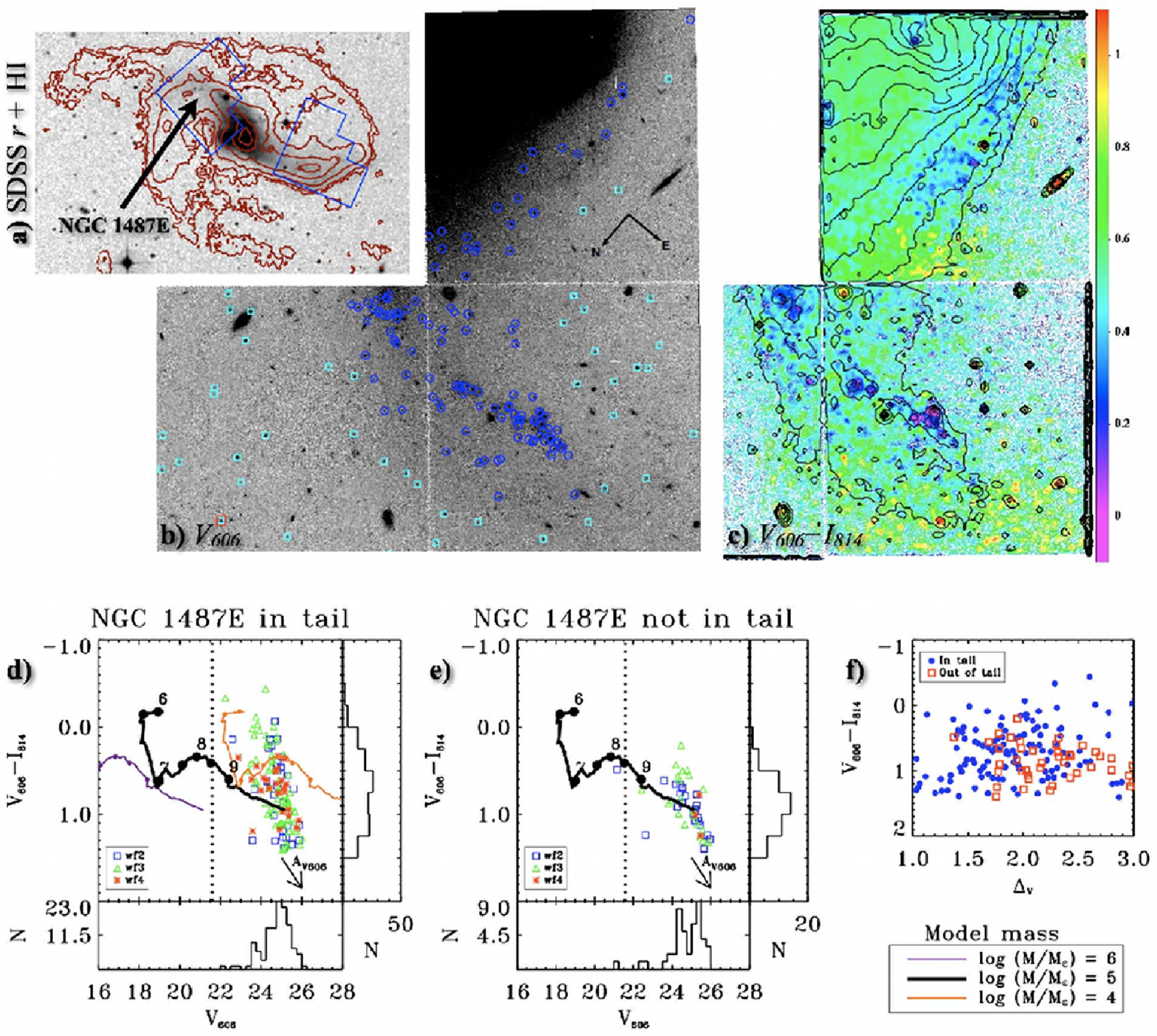}
\caption{\footnotesize{\textbf{NGC 1487E:} a) SDSS $r$-band image with WFPC2 footprint and N$_{\mbox{\scriptsize{\HI}}}$ contours in levels of 1, 2, 5, 10, 15, and 20 $\times $ 10$^{20}$ cm$^{-2}$ overlaid. b) \textit{HST} $V_{606}$-band mosaic, with in-tail sources indicated as blue circles and out-of-tail sources as cyan squares. Sources outlined in red are SCCs. c) \VI\ color map with \Vhst\ contours in units of -0.5 mag from the tail limit. d) CMD for in-tail sources. Different symbols correspond to different WFPC2 chips. The vertical dashed line indicates where M$_V$ = -8.5, transformed to WFPC2 instrumental magnitudes. Zero-age burst population synthesis models of \citet{BC03} (see text for details) are overlaid, with log mass (M/\msun) = 6, 5, and 4. Dots mark increases in log age from 6 to 9 in steps of 0.5. An extinction vector (1 mag) is labelled at the bottom right. e) Same as (d) but for out-of-tail sources. For clarity, only one evolutionary track of log mass (M/\msun) = 5 is given.  f) \VI--$\Delta_{V}$ diagram for all sources in-tail (blue circles) and out-of-tail (red squares). Full resolution Figures 3.1--3.17 for all tails are available in the electronic edition of {\it The Astrophysical Journal.}}}
\end{figure*}
\addtocounter{figcount}{1}


Population synthesis models of \citet{BC03} are overlaid in the CMDs. Evolutionary tracks are shown for a simple stellar population (SSP) given a zero-age burst of star formation, a \citet{salpeter} initial mass function, solar metallicity and total stellar mass of 10$^4$, 10$^5$, and 10$^6$ \msun. Numbers indicate the age of the burst (log yr), and one magnitude of extinction in \Vhst\ is shown as the labeled A$_{V606}$ vector. Estimating ages and masses of putative clusters will inevitably be complicated by aforementioned stochastic effects, metallicity-age-extinction degeneracies, and model inadequacies. We do not expect extinctions exceeding 1 mag, so internal extinction is partially mitigated compared to galaxy interiors. Thus, with all these uncertainties in mind, ages and masses of detected clusters must be treated qualitatively.

\subsection{``Beads on a String" Morphology} 

When smoothed with a 12-pixel ($\approx$ 1.2\arcsec) Gaussian kernel, three of the \VI\ color maps in the \mbox{Figure Set 3} show evidence for ``beads on a string" star formation, particularly for the bluest (and possibly youngest) objects we detect in the tails. NGC 1487E, MCG -03-13-063, and AM 1054-325 all exhibit this morphology when their spatial resolution is sufficiently downgraded, and our compact objects blend to clumps of $\sim$ 10$^5$ -- 10$^6$ \msun. These beads are all separated by $\sim$ 3 kpc, adjusting for possible tail inclinations.

We also compared these separations with those measured for a set of GALEX-imaged tidal tails presented in \citet{smith10}. These approaches are heuristically compatible, as both employ images with $>$ 1$\arcsec$ resolution and the identification of young blue clumps of star formation within them. In the GALEX sample, we remeasure clump separations of $\sim$ 1--7 kpc, with reasonable agreement between the GALEX-imaged beads of NGC 2535 (3--4 kpc; \citealp{smith10}) and its optical counterpart (our study).  

This type of star formation is typically attributed to gravitational instabilities within the host environment (\citealp{elmegreen83}; \citealp{bournaud04}), scaling with the local Jeans mass. Within spiral arms of galaxies, these clumps are separated by the Jeans wavelength (the minimum radius necessary for collapse), which for that environment is $\sim$ 1--4 kpc \citep{elmegreen83}. This is also similar to the clump separations observed in interacting galaxies at intermediate redshifts \citep{elm07}. The masses estimated here are at the low end of the observable mass range for the clumps of that study, which range from 10$^6$ to a few 10$^8$ \msun.

Star-forming clumps with separations $\sim$ 3 kpc, while consistent with gravitational collapse scales of spiral galaxies and interacting pairs at intermediate redshifts,  are not pronounced in all tails' color maps. It is possible a combination of metallicity, age/star formation history, orientation, and smoothing effects contribute to this observed morphology in some tails but not in others. There may be more physical motivations as well. This formation morphology is pronounced in small subsets of this sample and the tidal debris of interactions surveyed by \citet{smith10}, implying that not all tails share the physical conditions necessary for promoting these structures. Certain ISM velocity dispersions and/or mass densities within the host environment are required to produce these beads, but determining these local quantities in the cluster-forming regions is beyond the scope of this work.

\subsection{A Quantitative Comparison of SCCs} 

In this set of interacting pairs, several have visually striking abundances of in-tail sources relative to regions beyond the tail. To quantitatively compare these systems, the relative areas of in-tail and out-of-tail regions on the three chips must be considered, along with foreground and background object contamination by non-cluster objects. Distance-related selection effects must also be acknowledged. 

\mbox{Table 3} presents the tails, their host galaxy luminosity distances in Mpc (from the NASA Extragalactic Database), their completeness limits (\S 2.5) and their pixel sizes (in pc). We also include the number of SCCs detected within (N$_{\rm{in}}^{\rm{SCC}}$) and outside (N$_{\rm{out}}^{\rm{SCC}}$) of the debris, and the area of the in-tail and out-of-tail regions in kpc$^{2}$, A$_{\rm{in}}$, and A$_{\rm{out}}$. The in-tail excesses \SCC\ are then quantified as the difference between N$_{\rm{in}}^{\rm{SCC}}$/A$_{\rm{in}}$ and N$_{\rm{out}}^{\rm{SCC}}$/A$_{\rm{out}}$, with uncertainties from Poisson statistics of N$_{\rm{in}}^{\rm{SCC}}$ and N$_{\rm{out}}^{\rm{SCC}}$. Errors of 2.5$\sigma$ are recorded. If there are local clusters outside the optically-defined tail, this method for background subtraction is systematically too large and gives a lower limit to actual cluster surface density. Thus, SCC excesses at the 2.5$\sigma$ level are statistically significant.

 
 \begin{table*}[htbp]
\renewcommand{\thefootnote}{\alph{footnote}}
\begin{center}
\begin{tabular*}{1.0\textwidth}{@{\extracolsep{\fill}}lrrrrrrrrrr}
\multicolumn{11}{c}{\sc Table 3. In-Tail Cluster Candidate Excesses, M$_V$ $<$ -8.5} \\
\hline\hline
Tail & Distance & M$_{V, 50\%}$ & Pixel Size & N$_{\rm{in}}^{\rm{SCC}}$ & N$_{\rm{out}}^{\rm{SCC}}$ & A$_{\rm{in}}$ & A$_{\rm{out}}$ & N$_{\rm{in}}^{\rm{SCC}}$/A$_{\rm{in}}$ & N$_{\rm{out}}^{\rm{SCC}}$/A$_{\rm{out}}$ & $\Sigma_{{\rm SCC}}^a$\\
 	& (Mpc) & 	& (pc) & 	& 	& (kpc$^2$) & (kpc$^2$)	& (kpc$^{-2}$) & (kpc$^{-2}$)  &(kpc$^{-2}$)  \\
\hline
NGC 1487E & 10.8 & -4.6 & 5.22 & 0 & 1 & 16.49 & 40.93 & 0 & 0.024 & -0.024 $\pm$ 0.061 \\
NGC 1487W & 10.8 & -4.5 & 5.22 & 2 & 0 & 16.39 & 41.03 & 0.122 & 0 & 0.122 $\pm$ 0.216 \\
NGC 4747 & 20.2 & -6 & 9.75 & 7 & 5 & 43.66 & 68.01 & 0.160 & 0.074 & 0.087 $\pm$ 0.172 \\
NGC 520 & 27.2 & -6.4 & 13.13 & 0 & 4 & 26.40 & 375.97 & 0 & 0.011 & -0.011 $\pm$ 0.013 \\
NGC 2992 & 36.6 & -7.2 & 17.67 & 37 & 15 & 214.38 & 198.58 & 0.173 & 0.076 & 0.097 $\pm$ 0.086 \\
NGC 2993 & 36.6 & -7.3 & 17.67 & 15 & 7 & 142.10 & 270.86 & 0.106 & 0.026 & 0.080 $\pm$ 0.072 \\
NGC 2782E & 38.1 & -7.5 & 18.40 & 87 & 4 & 336.79 & 165.83 & 0.258 & 0.024 & 0.234 $\pm$ 0.076 \\
NGC 2782W & 38.1 & -7.3 & 18.40 & 10 & 7 & 235.91 & 266.71 & 0.042 & 0.026 & 0.016 $\pm$ 0.042 \\
MCG -03-13-063 & 46.2 & -7.8 & 22.31 & 88 & 26 & 230.18 & 632.29 & 0.382 & 0.041 & 0.341 $\pm$ 0.104 \\
ESO 376- G 028 & 52.9 & -8.1 & 25.54 & 4 & 60 & 80.97 & 1036.73 & 0.049 & 0.058 & -0.008 $\pm$ 0.065 \\
AM 1054-325 & 52.9 & -8.7 & 25.54 & 180 & 60 & 136.28 & 981.42 & 1.321 & 0.061 & 1.260 $\pm$ 0.247 \\
NGC 2444 & 58.2 & -8.2 & 28.10 & 27 & 15 & 716.94 & 543.56 & 0.038 & 0.028 & 0.010 $\pm$ 0.025 \\
NGC 2535 & 59.8 & -8.4 & 28.88 & 44 & 49 & 295.03 & 995.44 & 0.149 & 0.049 & 0.100 $\pm$ 0.059 \\
NGC 6872E & 62.6 & -8.3 & 30.23 & 163 & 63 & 588.99 & 999.53 & 0.277 & 0.063 & 0.214 $\pm$ 0.058 \\
NGC 6872W & 62.6 & -8.5 & 30.23 & 195 & 36 & 788.90 & 799.62 & 0.247 & 0.045 & 0.202 $\pm$ 0.048 \\
NGC 1614N & 65.6 & -8.6 & 31.68 & 33 & 28 & 362.63 & 1386.52 & 0.091 & 0.020 & 0.071 $\pm$ 0.041 \\
NGC 1614S & 65.6 & -8.6 & 31.68 & 43 & 28 & 288.50 & 1460.65 & 0.149 & 0.019 & 0.130 $\pm$ 0.058 \\
\hline
\multicolumn{11}{c}{The K03 Sample} \\
\hline
NGC 4038 & 13.8 & -5.1 & 6.66 &        3 &        3 & 76.92 & 76.92 & 0.039 & 0.039 & 0.000 $\pm$ 0.080 \\
NGC 3256E & 42.8 & -7.8 & 20.67 &       49 &       55 & 290.32 & 333.33 & 0.169 & 0.165 & 0.004 $\pm$ 0.082 \\
NGC 3256W & 42.8 & -7.8 & 20.67 &       62 &       58 & 205.88 & 428.57 & 0.301 & 0.135 & 0.166 $\pm$ 0.105 \\
NGC 7252E & 62.2 & -8.1 & 30.03 &       23 &       32 & 700 & 1181.82 & 0.033 & 0.027 & 0.006 $\pm$ 0.021 \\
NGC 7252W & 62.2 & -8.1 & 30.03 &       31 &       29 & 684.21 & 1250 & 0.045 & 0.023 & 0.022 $\pm$ 0.023 \\
NGC 3921S & 84.5 & -8.8 & 40.80 &       14 &       30 & 692.31 & 2333.33 & 0.020 & 0.013 & 0.007 $\pm$ 0.015 \\

\hline\hline
\end{tabular*} 
\\
\end{center}
$^a$Errors of 2.5$\sigma$ are quoted.

\end{table*}
\renewcommand{\thefootnote}{\arabic{footnote}}


Completeness may also have an effect on AM 1054-435 and NGC 1614N/S. Since the 50\% completeness limit of these systems is approximately the M$_V$ selection criterion (M$_{V,50\%}$ = -8.6 -- -8.7), these tails will have a systematic underrepresentation of in-tail and out-of-tail SCCs. We estimate, using a power-law LF with a logarithmic slope $\alpha$ = 2 -- 2.5 for the in-tail sources (see \S 3.3), and extrapolating from the corresponding \Vhst\ distribution for the out-of-tail sources, that the \SCC\ values quoted for these two tails may be underestimated by $\approx$ 25\%. No corrections to these excesses have been made. 

\mbox{Figure 4} outlines these results, including the interacting pairs of K03 to the right of the vertical dotted line, and their identically calculated excesses. Note that \SCC\ values for K03 tails are different than those originally published, because of our $V$-$I <$ 2.0 criterion and the different luminosity distances we use here (for consistency, H$_0$ = 73 km s$^{-1}$ Mpc$^{-1}$; we use the distance derived by \citealp{saviane} for NGC 4038). The horizontal line marks where \mbox{\SCC\ = 0}; any points within 2.5$\sigma$ of this level imply the host tail is most likely devoid of star clusters within the limit of our broadband observations (for clarity, only 1$\sigma$ errors are shown). Overplotted boxes indicate tails with statistically insignificant SCC excesses, including tails whose SCCs reside primarily in TDG candidates (NGC 2992 and NGC 7252E/W; see \S 3.2.2). Black circles show the results of the \SCC\ analysis performed with the \mbox{M$_V$ $<$ -8.5} and $V$-$I <$ 2.0 criteria. Blue triangles and red asterisks in \mbox{Figure 4} show the results of similar calculations, but with fainter M$_V$ limits by one and two magnitudes (M$_V$ $<$ -7.5 -- -6.5), respectively. As the M$_V$ threshold increases, NGC 1487 through NGC 520 enjoy increases in \SCC. More distant systems experience similar changes in SCC excess as the magnitude limit is increased, but only until their completeness limits are reached. 


\begin{figure*}[htbp]
\plotone{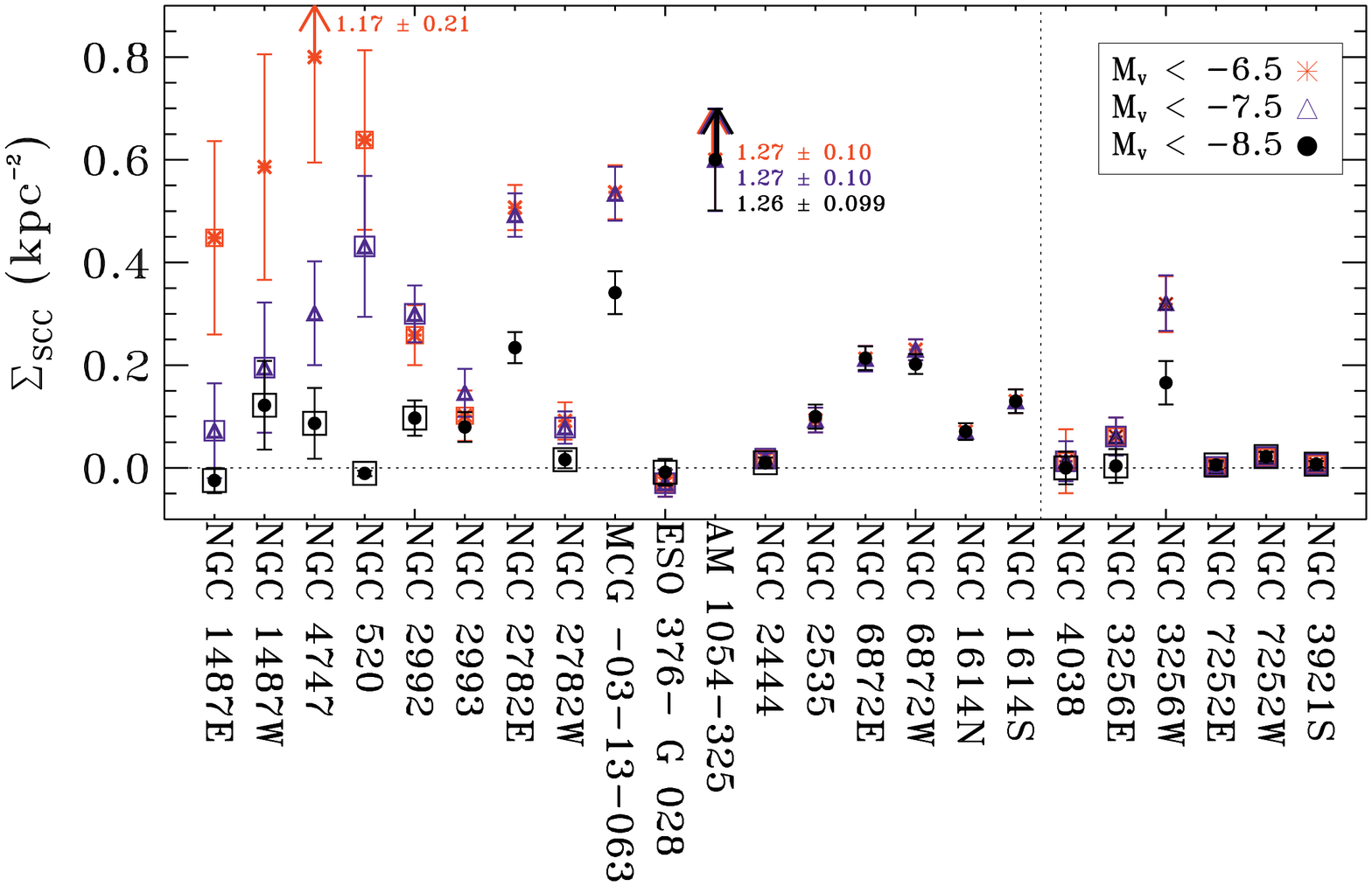}
\caption{Star cluster candidate density, \SCC\ (kpc$^{-2}$) for the tidal tail sample, given $V$-$I$ $<$ 2.0, and a M$_V$ cutoff of -8.5 (black circles), -7.5 (blue triangles), and -6.5 (red asterisks). Boxes indicate where \SCC\ is statistically insignificant, either directly from a SCC deficit, or from their biased location in TDG candidates. For clarity, only 1$\sigma$ errors are shown and several points for NGC 4747 and AM 1054-325 are plotted below their actual \SCC\ values. Systems left of the vertical dashed line are presented in this paper; tails to the right were investigated by K03.} 
\end{figure*}
 \addtocounter{figcount}{1}

To further explore the differences in source populations between in- and out-of-tail regions, we performed Kolmogorov-Smirnov (K--S) tests on the distributions of \Vhst\ and \VI\ for all in-tail vs.\ out-of-tail sources (N$_{\rm{in}}$, and N$_{\rm{out}}$); these are presented in \mbox{Table 4}, along with each tail's \SCC\ for comparison. The results of K03 are also shown. PKS$_{\mbox{\scriptsize{\Vhst}}}$ and PKS$_{\mbox{\scriptsize{\VI}}}$ values list the probabilities that  the in-tail and out-of-tail objects are drawn from the same distribution of \Vhst\ and \VI, respectively. This provides a measure of the independence of in-tail sources from the out-of-tail sources. Note that in-tail and out-of-tail populations are likely to be independent in terms of color and magnitude distributions if their K--S probabilities are $<$ 0.02, i.e.\ that the populations are distinct at the $\approx$ 2.5$\sigma$ level. In three cases (AM 1054-325 and NGC 1614N/S), tails with statistically distinguishable populations in color and magnitude also have statistically significant SCC populations. This largely confirms that these tails have clusters. Tails with \SCC\ = 0 all have PKS$_{\mbox{\scriptsize{\Vhst}}}$ and PKS$_{\mbox{\scriptsize{\VI}}}$ values$>$ 0.02, reaffirming that they are devoid of star clusters at the M$_V <$ -8.5 level.

Other tails with significant SCC excess but unfavorable K--S statistics still likely contain clusters, however. MCG -03-13-063, NGC 2535 and NGC 3256W have distinct source population differences in \VI\ and not in \Vhst, as well as SCC excesses. The similarities in the magnitude distributions may be due to a slight contamination of foreground sources for tails at low galactic latitudes ($|b| <$ 30$^{\circ}$. in these cases). NGC 2993 and NGC 6872E/W conversely show significant source distribution differences in magnitude but not in color. For NGC 2993, its \HI\ tail extends beyond its optical debris, so contamination of the out-of-tail sources by faint, similarly-aged clusters within the \HI\ tail but beyond the optical debris may account for its high PKS$_{\mbox{\scriptsize{\VI}}}$. Sources within NGC 6872E/W debris may have variable extinction and metallicities \citep{iraklis09}, and/or may be contaminated by red (\VI\ $\lesssim$ 1) foreground sources ($b \approx$ -32.5$^{\circ}$). Lastly, NGC 2782E has a nonzero SCC excess, but shows insignificant source differences between its in-tail and out-of-tail regions in both \Vhst\ and \VI. It is probable that the out-of-tail regions of NGC 2782E are undersampled --  with only 8 sources detected, it is difficult to determine their independence in color and magnitude from the ample in-tail objects.

In summary, by combining K03 with this study, cluster candidate excesses at M$_V <$ -8.5 exist at the 2.5$\sigma$ level in 10 out of 23 tails (NGC 2993, NGC 2782E, MCG -03-13-063, AM 1054-325, NGC 2535, NGC 6872E/W, and NGC 1614N/S, and NGC 3256W). Our detection of these SCCs in the tidal debris regions are all complete to at least $\approx$ 50\% to that magnitude. Many of the tails with excesses also have \VI\ and \Vhst\ distributions that differ between tail and out of tail sources, indicating the in tail sources are real and associated with the tail. The tails with statistically nonzero cluster candidate excesses and source distributions indistinguishable from out-of-tail objects in color or magnitude can be explained by foreground contamination or poor statistics from undersampled out-of-tail areas. These tails are still likely to have clusters.


\begin{table*}[htbp]
\renewcommand{\thefootnote}{\alph{footnote}}
\begin{center}
\begin{tabular*}{1.0\textwidth}{@{\extracolsep{\fill}}lrrrr|r}

\multicolumn{6}{c}{\sc Table 4. K--S probabilities} \\
\hline\hline
Tail &  N$_{\rm{in}}$ & N$_{\rm{out}}$ & PKS$_{\mbox{\scriptsize{\Vhst}}}$ &  PKS$_{\mbox{\scriptsize{\VI}}}$ & \SCC$^a$ (kpc$^{-2}$) \\
\hline
NGC 1487E &      122 &       45 & 0.341 & 0.0263 & -0.024 $\pm$ 0.061 \\
NGC 1487W &      121 &       67 & 0.593 & 0.396 & 0.122 $\pm$ 0.216 \\
NGC 4747 &      117 &       45 & 0.878 & 0.646 & 0.087 $\pm$ 0.172 \\
NGC 520 &       21 &       66 & 1.37 $\times$ 10$^{-3}$ & 0.757 & -0.011 $\pm$ 0.013 \\
NGC 2992 &      104 &       45 & 0.0187 & 0.660 & 0.097 $\pm$ 0.086 \\
NGC 2993 &       37 &       43 & 0.0172 & 0.652 & 0.080 $\pm$ 0.072 \\
NGC 2782E &      187 &        8 & 0.875 & 0.963 & 0.234 $\pm$ 0.076 \\
NGC 2782W &       49 &       31 & 0.0588 & 0.249 & 0.016 $\pm$ 0.042 \\
MCG -03-13-063 &      141 &       48 & 0.324 & 0.0617 & 0.341 $\pm$ 0.104 \\
ESO 376- G 028 &        4 &       82 & 0.793 & 0.0735 & -0.008 $\pm$ 0.065 \\
AM 1054-325 &      184 &       82 & 2.36 $\times$ 10$^{-10}$ & 9.22 $\times$ 10$^{-25}$ & 1.260 $\pm$ 0.247 \\
NGC 2444 &       34 &       16 & 0.545 & 0.241 & 0.010 $\pm$ 0.025 \\
NGC 2535 &       45 &       59 & 0.0904 & 1.16 $\times$ 10$^{-4}$ & 0.100 $\pm$ 0.059 \\
NGC 6872E &      180 &       93 & 7.26 $\times$ 10$^{-5}$ & 0.209 & 0.214 $\pm$ 0.058 \\
NGC 6872W &      219 &       38 & 5.17 $\times$ 10$^{-3}$ & 0.119 & 0.202 $\pm$ 0.048 \\
NGC 1614N &       33 &       28 & 1.94 $\times$ 10$^{-3}$ & 3.33 $\times$ 10$^{-4}$ & 0.071 $\pm$ 0.041 \\
NGC 1614S &       43 &       28 & 2.61 $\times$ 10$^{-4}$ & 1.27 $\times$ 10$^{-6}$ & 0.130 $\pm$ 0.058 \\
\hline
\multicolumn{6}{c}{The K03 Sample} \\
\hline
NGC 4038 &       44 &       26 & 0.270 & 0.890 & 0 $\pm$ 0.080 \\
NGC 3256E &       70 &       60 & 0.200 & 1.00 $\times$ 10$^{-3}$ & 0.004 $\pm$ 0.082 \\
NGC 3256W &      106 &       84 & 0.0300 & 2.20 $\times$ 10$^{-6}$ & 0.166 $\pm$ 0.105 \\
NGC 7252E &       25 &       40 & 0.950 & 0.910 & 0.006 $\pm$ 0.021 \\
NGC 7252W &       33 &       34 & 0.0500 & 0.900 & 0.022 $\pm$ 0.023 \\
NGC 3921S &       14 &       30 & 0.980 & 0.0500 & 0.007 $\pm$ 0.015 \\
\hline\hline
\end{tabular*} 
\\
\end{center}
$^a$Errors of 2.5$\sigma$ are quoted.

\renewcommand{\thefootnote}{\arabic{footnote}}
\end{table*}


\subsubsection{\SCC\ and Global Tail Characteristics}  

At the M$_V$ $<$ -8.5 limit, several tails in our sample have comparable/larger SCC excesses to that of NGC 3256W in K03. NGC 6872E/W for example has an average excess of $\approx$ 0.2 kpc$^{-2}$, compared to $\approx$ 0.16 kpc$^{-2}$ for NGC 3256W. According to \mbox{Table 2}, both of these systems have long tidal tails that are rich in \HI\ (\mhi\ $\gtrsim$ 2 $\times$ 10$^{9}$ \msun), but have drastically different merging mass ratios. The tidal debris of MCG -03-13-063 and AM 1054-325 have drastically higher SCC densities ($\approx$ 0.34 and 1.3 kpc$^{-2}$, respectively), and are both younger mergers by $\approx$ 300 Myr. They also have different merging mass ratios, and their total \HI\ masses are unknown.  NGC 1614S has a close correspondence of cluster excess ($\approx$ 0.13 kpc$^{-2}$) to NGC 3256W and it is a product of a major merger, but it is a low \mhi, previously merged system with debris $>$ 200 Myr older than NGC 3256. 

It appears that the surface density of clusters depends on the complex interplay of the global tail parameters of \mbox{Table 2}, if they are are indeed important. That is not to say there are no trends in the combined samples at all, but disentangling statistical influences of single parameters is difficult for this finite sample size. For instance, there may be an overall preference for cluster formation at young interaction ages, as the highest \SCC\ values belong to tails younger than 250 Myr. However, K--S tests demonstrate that tails with significant SCC excesses and tails without them have a probability of 0.715 of sharing a common age distribution. Age alone apparently does not direct cluster formation. Combined with other parameters like WFPC2 tail surface brightness and color, age may still have a role; this is addressed in \S 4.1.  
  
Furthermore, minor mergers appear as well suited for star cluster formation as major mergers; approximately 60\% of either type of merger has a nonzero SCC density for M$_V <$ -8.5. K--S tests reveal a probability of 0.608 that tails with both statistically zero and nonzero SCC excesses are drawn from the same distribution of interacting mass ratios. By its qualitative nature, the distribution of mass ratios is very coarse, so such a high value is not unexpected.
  
In terms of host galaxy SFR, no outstanding distinction is seen for cluster-abundant tails. From the discussion of individual systems in Appendix A, it is unlikely that these host galaxy SFRs correlate to \SCC\ in a systematic way, owing to different host galaxy orientations and/or the localized nature of star formation \citep{bergvall03}. K--S tests performed on the combined SCC-harboring and SCC-deficient samples also indicate no difference between these two tail populations (we find a probability of 0.787). This is in opposition to K03, who speculated on a connection between the SFR of NGC 3256 (33.4 \msun\ yr$^{-1}$, the highest of their selection of tails) and the cluster candidate excess observed for its western tail. NGC 1614 has a similarly high IR-derived SFR ($\approx$ 35 \msun\ yr$^{-1}$), but is superceded by other systems in the sample with larger SCC excesses in their tails and lower \textit{IRAS}-derived SFRs, e.g.\ AM 1054-325 and NGC 6872E/W. Projection effects and ULIRGS in our sample are probable culprits for this behavior. 

Likewise, most tail lengths cannot be accurately de-projected from their measurements in \mbox{Table 2}, so comparing \SCC\ to them is prohibitively qualitative. For \mhi, we calculate K--S probabilities of 0.886 for the SCC-differentiated tail samples. As in K03, overall gas richness is not an immediately crucial characteristic for forming massive star clusters.
 
In short, the solution set of global properties from the present dataset that are linked to an SCC density at M$_V <$ -8.5 are not unique. There appears to be no single global property that universally regulates a tail's capacity for forming star clusters at the M$_V <$ -8.5 level. We address the potential importance of WFPC2 FOV-scale properties in \S 4.  

\subsubsection{\SCC\ and TDG candidates}  

K03 suggested that the formation of tidal dwarf galaxies and star clusters in tidal tails are mutually exclusive. In our sample, the optical condensation present in the tidal tail of NGC 2992 is likely a true TDG \citep{bournaud04}, so further care must be taken to disentangle the tail sources from those attributed to this object. High resolution WFPC2 images of NGC 2992 in \mbox{Figure 4.5} reveal more structure within what may be a smaller TDG than ground-based observations would indicate (contrast the SDSS image with the \Vhst-band mosaic of that tail). Ignoring the SCCs attributable to the TDG candidate, we count N$_{in}^{SCC}$ = 17. Replacing the previous value (N$_{in}^{SCC}$ = 25) and subtracting an estimate of the TDG area ($\sim$ 10--20 kpc$^2$) from A$_{in}$ in Table 4, we find \SCC\ $\approx$ 0.01 $\pm$ 0.07 kpc$^{-2}$. This measurement is obviously not significant at our required 2.5$\sigma$ level. At M$_V <$ -7.5 and M$_V <$ -6.5, \SCC\ is still statistically insignificant; we calculate values $<$ 0.1 $\pm$ 0.3 for both cases.  The high inclination of NGC 2992 may also artificially inflate \SCC\ by providing a smaller projected tail area, so this must be interpreted even more conservatively.  

Furthermore, while our NGC 520 pointing focuses on the TDG candidate at the tip of the galaxy's tail (see Appendix A.4 for an alternative explanation), ACS images for NGC 520 at similar depths show evidence for clusters at the base of the tail (Chandar et al.\ 2011, in preparation). Although our SCCs in Figure 3 and in later Figures are flagged as belonging to the TDG candidate, star clusters likely exist in this system's tidal tail. Without deep imaging of the tail's optical extent, however, it is unknown whether they exist beyond the tail base. For now, this system may be seen as a tentative counterexample.   

NGC 2782 makes a stronger case against TDG-tail cluster candidate mutual exclusivity. There is kinematic evidence for a tidal dwarf in NGC 2782E \citep{yoshida94}; this is visible as an optical condensation in the upper left corner in chip WF3 in \mbox{Figure 4.7}). This kpc-sized agglomeration has blue colors (\VI\ $\sim$ 0.4) and emission line properties ([OIII], H$\alpha$+ [NII], and [SII]) consistent with a $\sim$ 10$^8$ \msun, metal-poor \HII\ region. However, it contains few SCCs and the rest of the tail retains a substantial cluster candidate excess beyond the 2.5$\sigma$ level (0.196 $\pm$ 0.07 kpc$^{-2}$). 

Moreover, AM 1054-325 has both a TDG candidate \citep{Weilbacher} and the highest \SCC\ of the tails in this sample. Unfortunately, there are no high-resolution kinematic data present to evaluate rotations or masses for its TDG candidate and confirm or deny TDG status, so this inference is not completely proven. No TDG-masked calculations of \SCC\ have been performed for this system, but even an in-tail SCC decrease of a factor of $\sim$ 5 would leave a statistically significant \SCC\ ($\approx$ 0.2 $\pm$ 0.11 kpc$^{-2}$). 

Though not absolute, there do seem to be counterexamples of the exclusivity of SCC and TDG formation in tidal tails. TDG formation appears favorable in tails with high velocity dispersions, and high Jeans masses \citep{elm93}. These objects may preferentially arise from interactions with extended dark matter halos, which allow efficient gas removal from host disks \citep{duc04} and may prevent strong episodes of cluster formation along the tails. Young tails with TDG candidates and SCCs may simply be still gas-rich enough along the debris for measurable SCC formation to occur. This cannot be confirmed without additional examples of these types of tails, with concordant \HI\ or CO observations.

\subsection{A Statistical Diagnostic of Tail Cluster Populations} 

The variations in SCC quantities we see in the tail sample and for different magnitude limits must depend of the logistics of star cluster \textit{formation} in tidal tails, as well as the details of their \textit{evolution}. Each generation of star clusters is imprinted with the physical mechanisms that dictate how they form. What we observe forms the basis for constructing the clusters' luminosity functions (LFs), which encode their underlying, heavily age-dependent mass functions (MFs). These MFs are shaped over time from cluster initial mass functions (CIMFs) by conditions in the tidal tail environment and their own internal physics.

One method to tackle the LF is to plot the the brightest cluster M$_V$ vs.\ number of clusters observed brighter than M$_V$ = -9 in each tail. Linear relationships between these two varaibles arise if the luminosities of the clusters are drawn from a power-law distribution LF with a slope $\approx$ -2.0 (\citealp{whitmore07}; \citealp{weidner04}; \citealp{billett02}). This size-of-sample effect connects a heightened local star formation rate (introducing large numbers of clusters) with an increasing probability of producing high-mass, luminous clusters (brightest M$_V$) In effect, the actual LF may not be a uniform power law, but appears as one for particular luminosity ranges. For the luminosities spanned by our SCC populations, \citet{gieles09} suggests that a more accurate LF slope is $\approx$ -2.5, while \citet{larsen09} offers $\approx$ -2.2 for populations of more strongly disrupting clusters. These values are strictly for spiral galaxies.  

We present our results in \mbox{Figure 5} for two test cases -- one where we count all in-tail SCCs in defining the number of clusters (Figure 5a), and a second where we statistically subtract out-of-tail SCCs to estimate the cluster number (Figure 5b). The latter method subtracts the number of out-of-tail SCCs per unit area (N$_{\rm{out}}^{\rm{SCC}}$/A$_{\rm{out}}$) from the number density within the tail (N$_{\rm{in}}^{\rm{SCC}}$/A$_{\rm{in}}$) and multiplies that by the area of the tail A$_{\rm{in}}$. The ``true" number of clusters should fall between these bracketing situations. In both cases, NGC 3256E/W and NGC 7252E/W have been removed, as these tails have unreliable, anomalously luminous brightest SCCs (M$_V \lesssim$ -14) that are not likely clusters. These tails exhibit a degree of contamination in their CMDs (K03) that prevents a safe determination of this quantity.


\begin{figure}[htbp]
\centering

\includegraphics[width=0.40\textwidth]{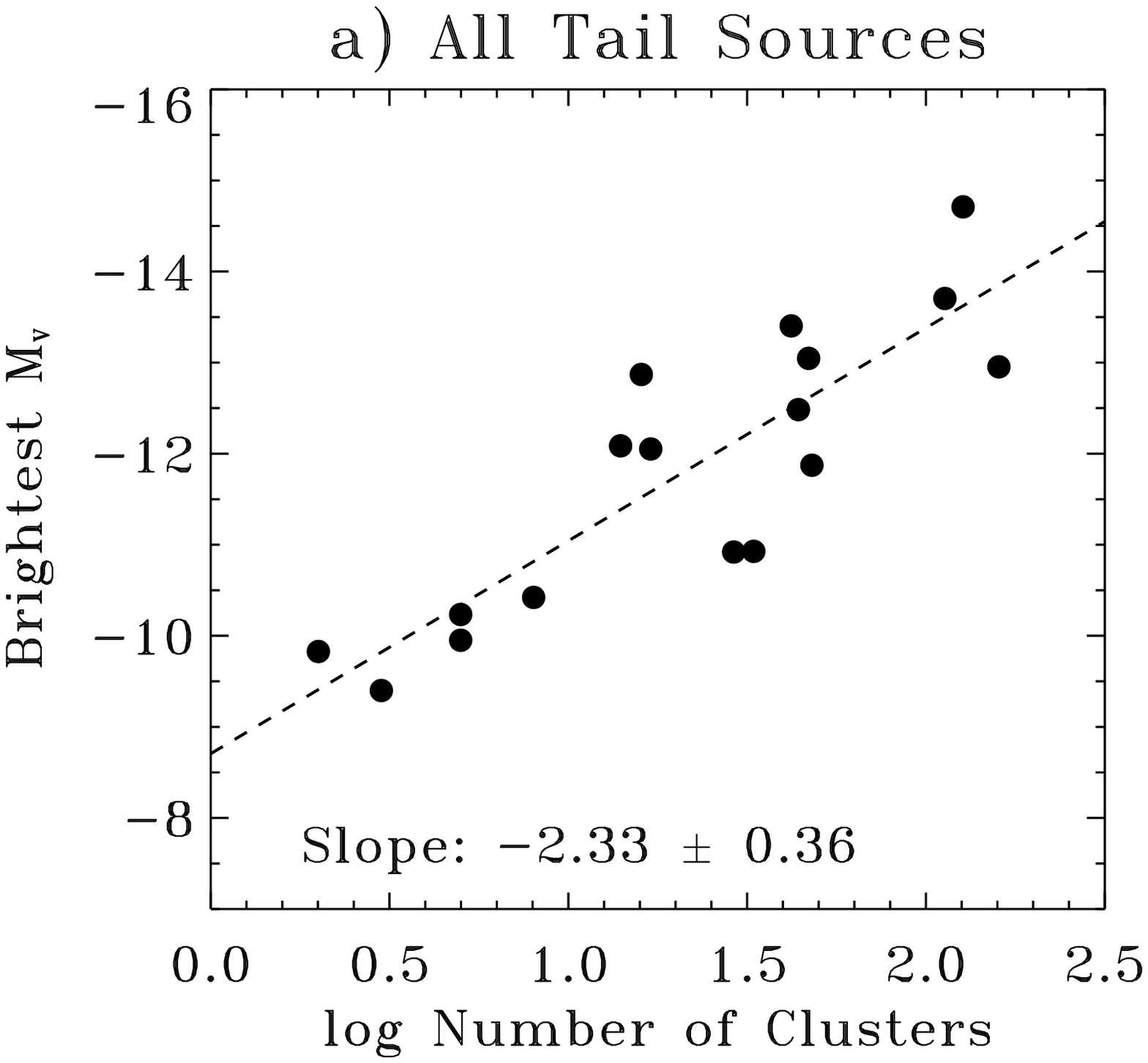}
\includegraphics[width=0.40\textwidth]{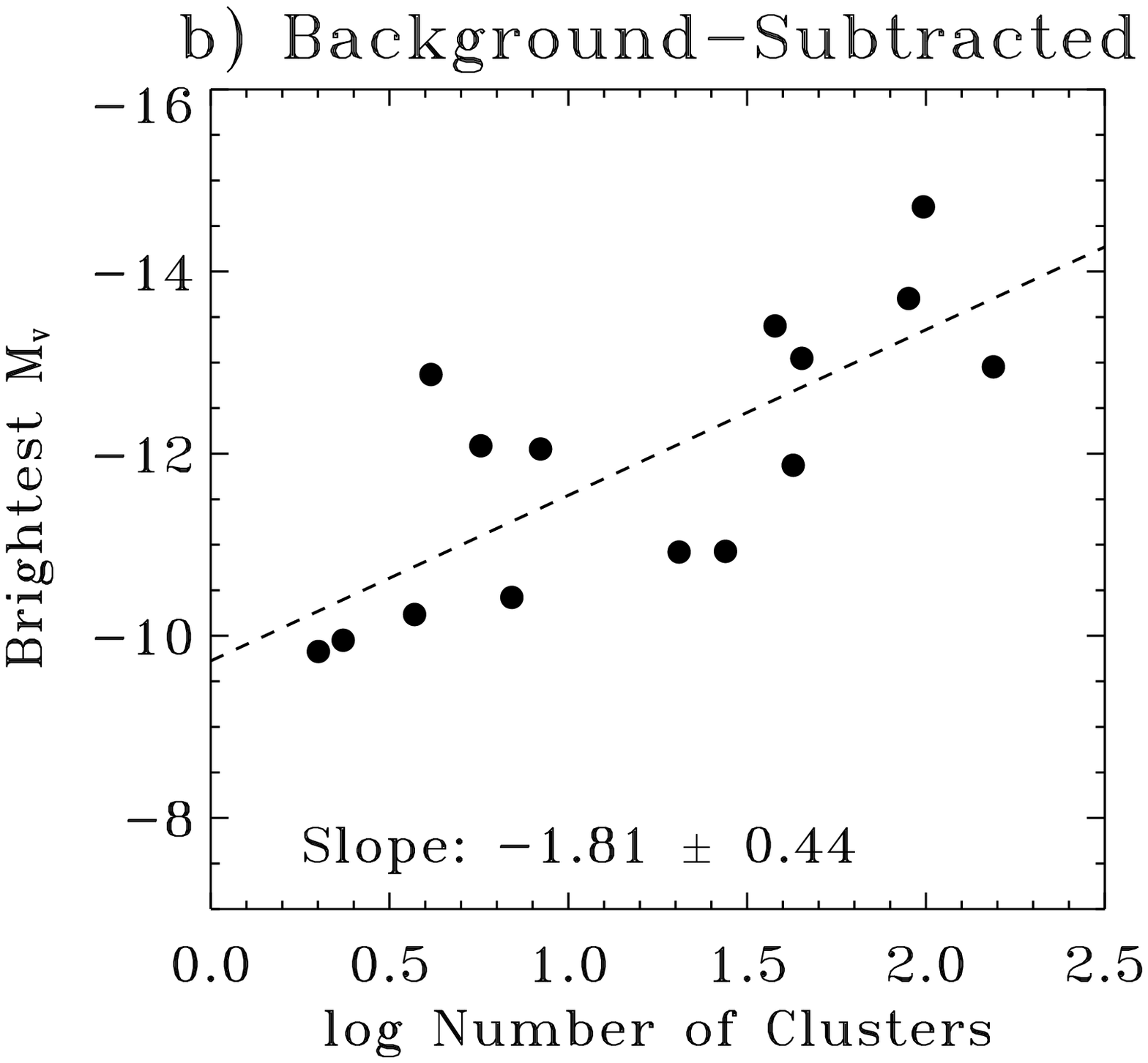}

\caption{Brightest M$_V$ -- log N plots for the in-tail SCC subset satisfying M$_V <$ -9 for all tidal debris  (a), and the background-subtracted case (b). In both cases, only tails with nonzero numbers of SCCs and reliable brightest magnitudes (i.e.\ high galactic latitudes) were used. It is possible the slopes of the actual linear trend between these two are consistent with the value -2.3 $\pm$ 0.2 observed for cluster populations in local starburst and quiescent galaxies (e.g.\ \citealp{whitmore07}).} 
\end{figure}
 \addtocounter{figcount}{1}


The slope we obtain for the resulting trend is -2.33 $\pm$ 0.36 for the first case, and -1.81 $\pm$ 0.44 for the second. These are in rough agreement within errors with the value -2.3 $\pm$ 0.2 derived from clusters in a variety of starbursts and quiescent spirals by \citet{whitmore07}. Our uncertainties are higher given the nature of this study, but ultimately this indicates that the underlying luminosity functions of clusters in our tidal tail sample are similar to those elsewhere in the local universe. The apparent similarity in LFs between our tidal tail clusters and elsewhere in the local universe is insightful. For LF slopes $\approx$ -2 -- -2.5, this implies that the underlying mechanisms at work in tidal tails are extremely similar in effect, if not in form, to those shaping the cluster populations observed in starburst and quiescent galaxies. Of course, our picture is incomplete, as star cluster LFs in tidal debris may be complicated by complex formation and dissolution histories from multiple sequences of extensive and compressive tides \citep{renaud09}, and/or their passage through the ISM (e.g.\ \citealp{gieles06}). There will also be stochastic effects at detectable masses and contributions from clusters born at different ages, as well as systematic effects and unseen variables in this study, all of which may contribute to the trends or scatter witnessed in \mbox{Figure 5}.



\section{Discussion}

\subsection{SCC Excess, Interaction Age, and Optical Tail Characteristics}

Though the dependence of \SCC\ on the properties in \mbox{Table 2} is generally not straightforward,  the most SCC-rich tails are all younger than 250 Myr. Since the \SCC\ differences cannot be statistically attributed to age alone, there may be other factors at work. To elucidate this, we plot a tail $V$ vs.\ $V$-$I$ CMD in \mbox{Figure 6} (panel a), along with \SCC\ vs.\ tail $V$ (panel b). Each graph presents the combined tail sample of this work and K03, color-coded by the progenitor system's interaction age. Boxed data points again indicate tails with no evidence of SCCs at the M$_V$= -8.5 level. An extinction vector of 1 magnitude in $V$ is shown in (a), and a histogram of the surface brightnesses of cluster-rich and cluster-empty tails is displayed in (b). We also note that NGC 2444 had no optical tail and was eliminated from this Figure.


\begin{figure}[htbp]
\centering
\begin{tabular}{c}
\epsfig{file=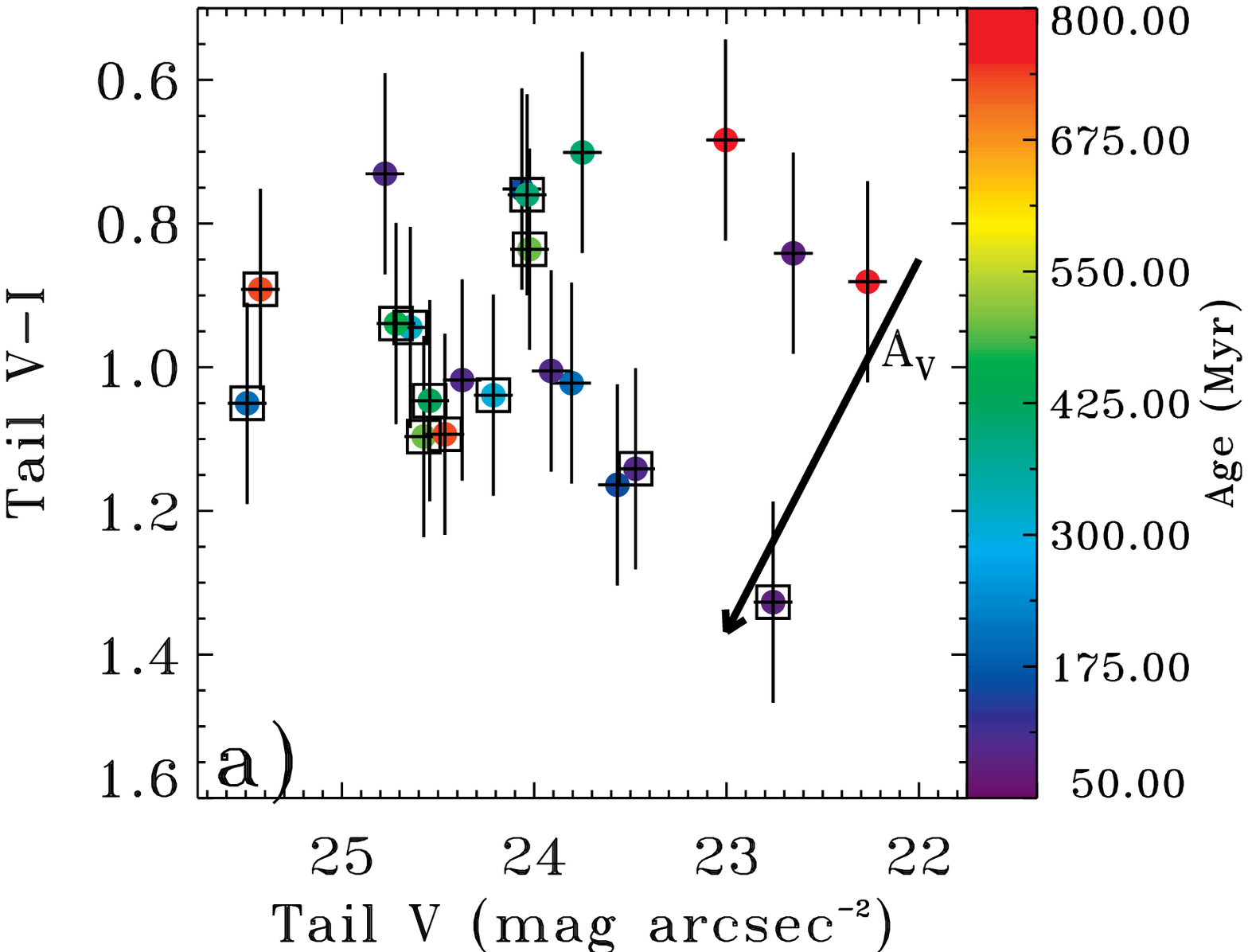,width=0.95\linewidth,clip=0}  \\
\epsfig{file=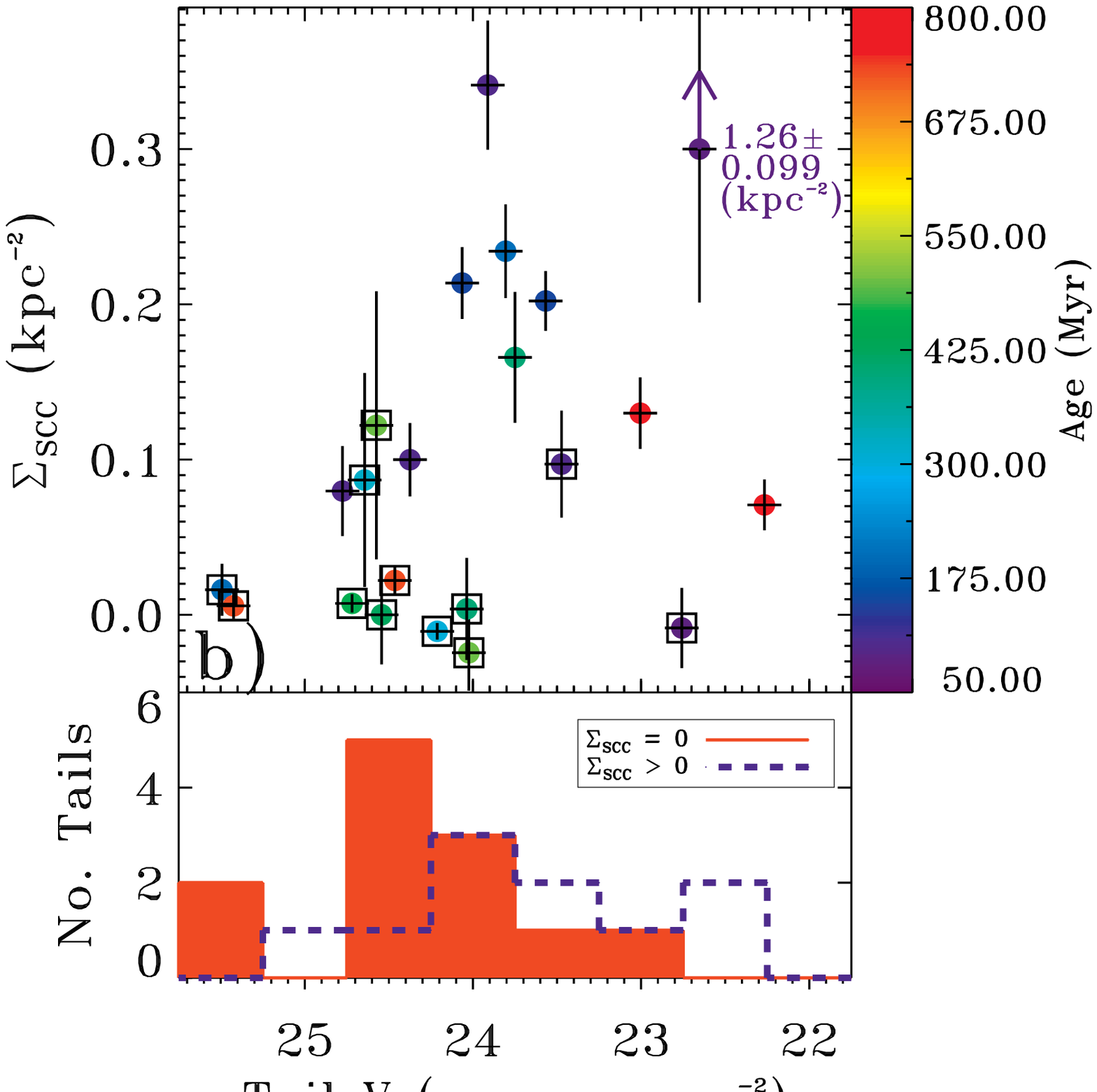,width=0.95\linewidth,clip=0} \\

\end{tabular}
\caption{The relationship between \SCC, interaction age, and tail optical properties. a) The color-magnitude distribution of the combined tidal tail sample, with colored points corresponding to the tails' interaction ages. Boxes indicate tails statistically devoid of SCCs, and all error bars show 1$\sigma$ uncertainties as in \mbox{Figure 5}. b) \SCC\ vs.\ tail $V$, highlighting tail ages. The histogram beneath this plot evinces the possibly distinct populations of tail surface brightnesses that constitute the tails with nonzero \SCC\ excesses (blue dashed lines) and those statistically without cluster candidates (solid red distribution).} 
\end{figure}
 \addtocounter{figcount}{1}

\mbox{Figure 6a} shows that clusters are found in tails of a variety of surface brightness, $V$-$I$ colors, and ages. As was seen in \S3.2.1, the tail sample is simply too phenomenologically diverse for distinguishable tail populations in terms of age. For example, NGC 2992 and NGC 2782W are young tails devoid of M$_V <$ -8.5 clusters while the $\approx$ 750 Myr old debris of NGC 1614N/S have two of the highest SCC excess of the sample (\SCC\ $\sim$ 0.1 kpc$^{-2}$). 

In terms of color, multiwavelength observations of other tidal tails \citep{smith10} that better circumvent the metallicity-extinction degeneracy show that blue colors typically imply young star formation and/or subsolar metallicity, rather than extinction effects. \citet{boquien09} provide a similar perspective on extinction. Thus, one might expect blue tail colors to correlate with clustered star formation here. However, K--S tests reveal that tails with \SCC\ statistically consistent with zero have a 33.1\% probability of sharing the same $V$-$I$ distribution as those with \SCC\ $>$ 0. There may be a number of explanations for this behavior. For instance, small variations of even 0.5 magnitudes of internal extinction may erase any trends of \SCC\ with tail $V$-$I$, which may be already unclear with our finite, 23-tail sample size. It is also possible that the blue colors expected of star formation are localized phenomena in most moderately star-forming tails, much as it is inferred in \citet{bergvall03} for galaxy interiors.  

Comparing \SCC\ excesses to tail surface brightness, we find a K--S probability of 0.0102 that cluster-bearing and cluster-devoid tails belong to the same tail $V$ distribution. Evidently, tails with cluster excesses tend to be brighter. This is presented in \mbox{Figure 6b} to demonstrate that it is not simply a trivial relationship of having more luminous clusters contributing to an overall more luminous tail. The accompanying histogram reflects the statistically significant difference between cluster-rich and cluster-poor tails in terms of surface brightness, but it is also a reminder of the small sample size and the difficulties in interpretation, especially given variations in internal extinctions and projection effects.

This could still be a promising result. Drawing upon the work of \citet{br06} and \citet{elmegreen93}, \citet{meurer09} reason that the R-band surface brightness roughly traces the midplane pressure of galaxies. This pressure regulates the pressure within molecular clouds, which is important for producing clustered star formation and other bound structures \citep{elm97}. Although many of the underlying assumptions of this deduction -- hydrostatic equilibrium of thin gas layers embedded in thin stellar disks, and the relative contributions of both of these components and their dynamics -- are invalid for tidal tails, the general preference for cluster-forming tails to be brighter on average in $V$ (which of course is not the same as $R$) may present a parallel paradigm. On more local scales tail surface brightness may offer a more accurate diagnostic of ambient pressure and the star clusters that form, but it is likely that the contribution of the gas potential and dynamics also needs to be more thoroughly addressed.      

Interestingly, the most cluster-rich tails (\SCC\ $\gtrsim$ 0.2 kpc$^{-2}$) are all young ($\lesssim$ 250 Myr), and brighter than the median ($V$ $\lesssim$ 24 mag arcsec$^{-2}$). Though we determined the age distributions for cluster-rich and cluster-devoid tails were indistinguishable, interaction age may be an additional ``bonus" for tails like AM 1054-325, MCG -03-13-063, NGC 2782E, and NGC 6872E, whose images capture strong bursts of star formation relatively near periapse for their interactions. This is typically seen in models of star-forming tidal debris (\citealp{chien10}; \citealp{dim08}, \citealp{m&h94}). Thus, should higher surface brightness loosely trace higher ambient pressures and thus star formation capacity, SCC detection is contingent on strong overall star formation, which is typically seen at younger dynamical ages for a variety of interacting systems.  

\subsection{SCC Excess and \HI}

Although in \S 3.2 there was no clear relationship between tail gas richness and cluster candidate excess, the \HI\ content of tidal tails warrant further investigation. In \mbox{Figure 7} we explore the complex relationship between the optical/\HI\ properties of the tidal tails and their \SCC\ measurements. Boxed points are the same as in the previous Figures, and we plot points whose ``color-coded" data are either missing or unreliable in black. This includes the abnormally high \HI\ densities of the NGC 1487E/W tails, which exhibit a superposition of disk and tail gas (\HIdens\ $>$ 10$^7$ \msun\ kpc$^{-2}$), and the brightest cluster candidate magnitudes of NGC 3256E/W and NGC 7252E/W (\S 3.3).


\begin{figure*}[htbp]
\centering
\begin{tabular}{cc}
\epsfig{file=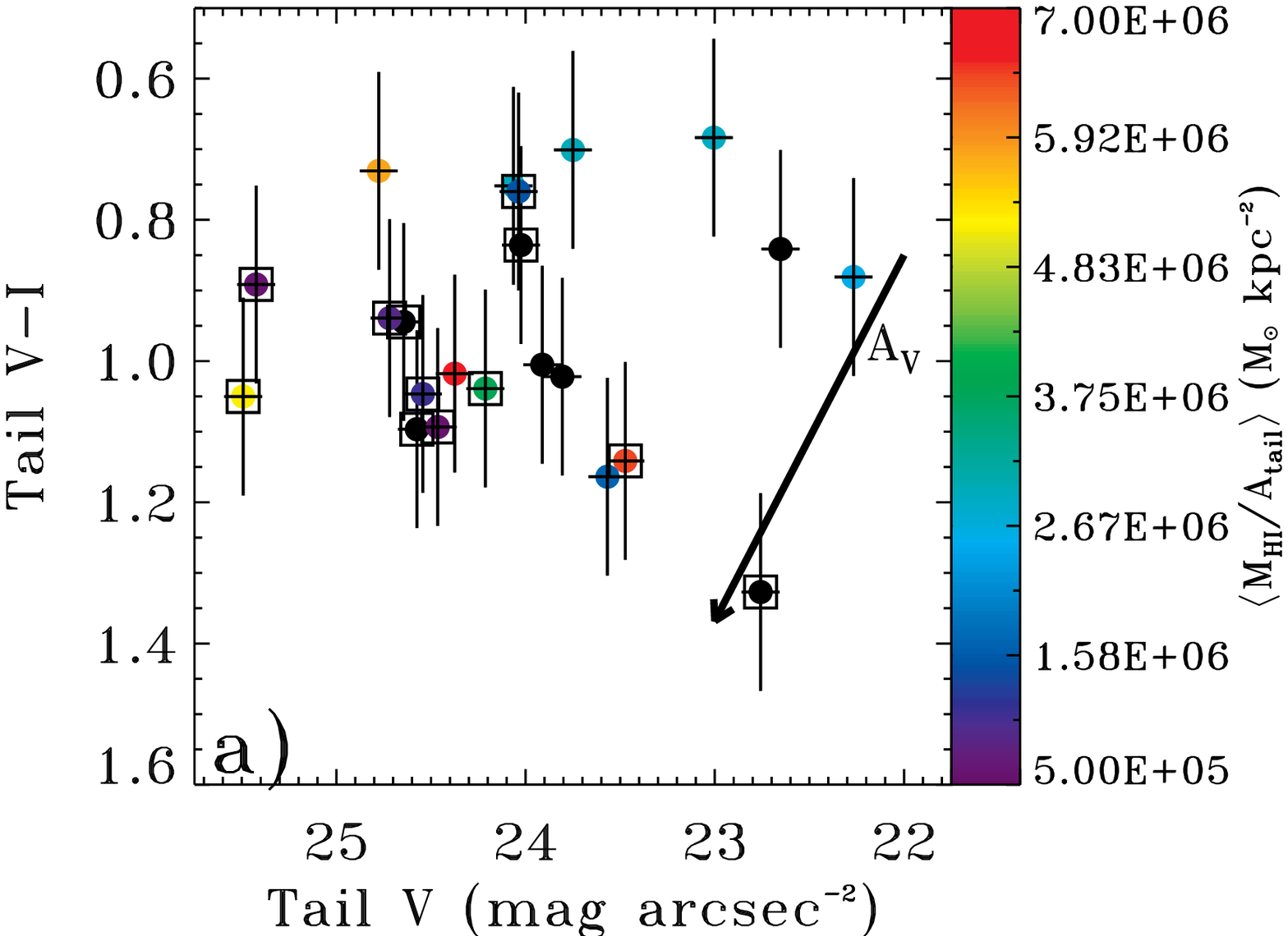,width=0.45\linewidth,clip=0}  &
\epsfig{file=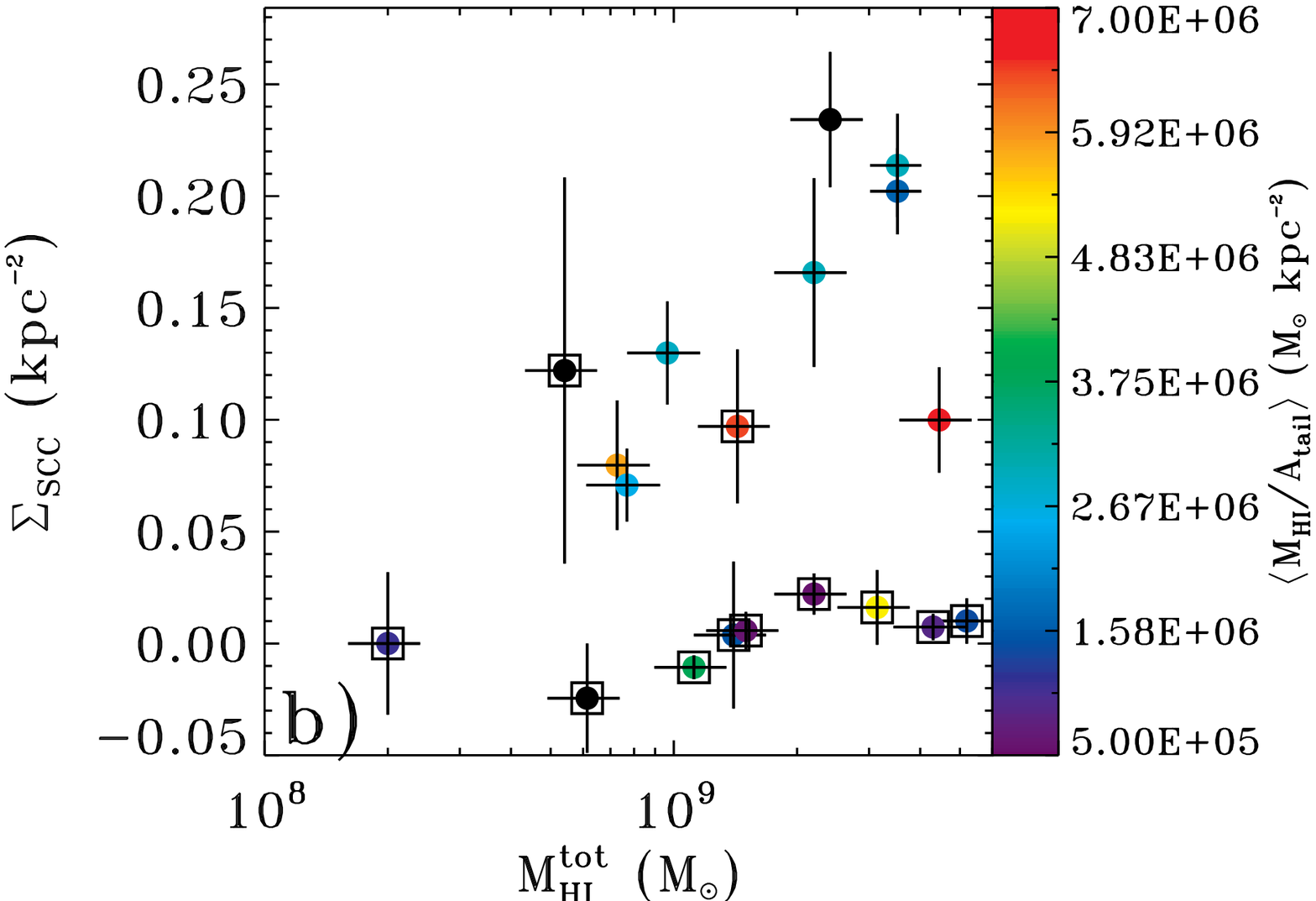,width=0.45\linewidth,clip=0}  \\
\epsfig{file=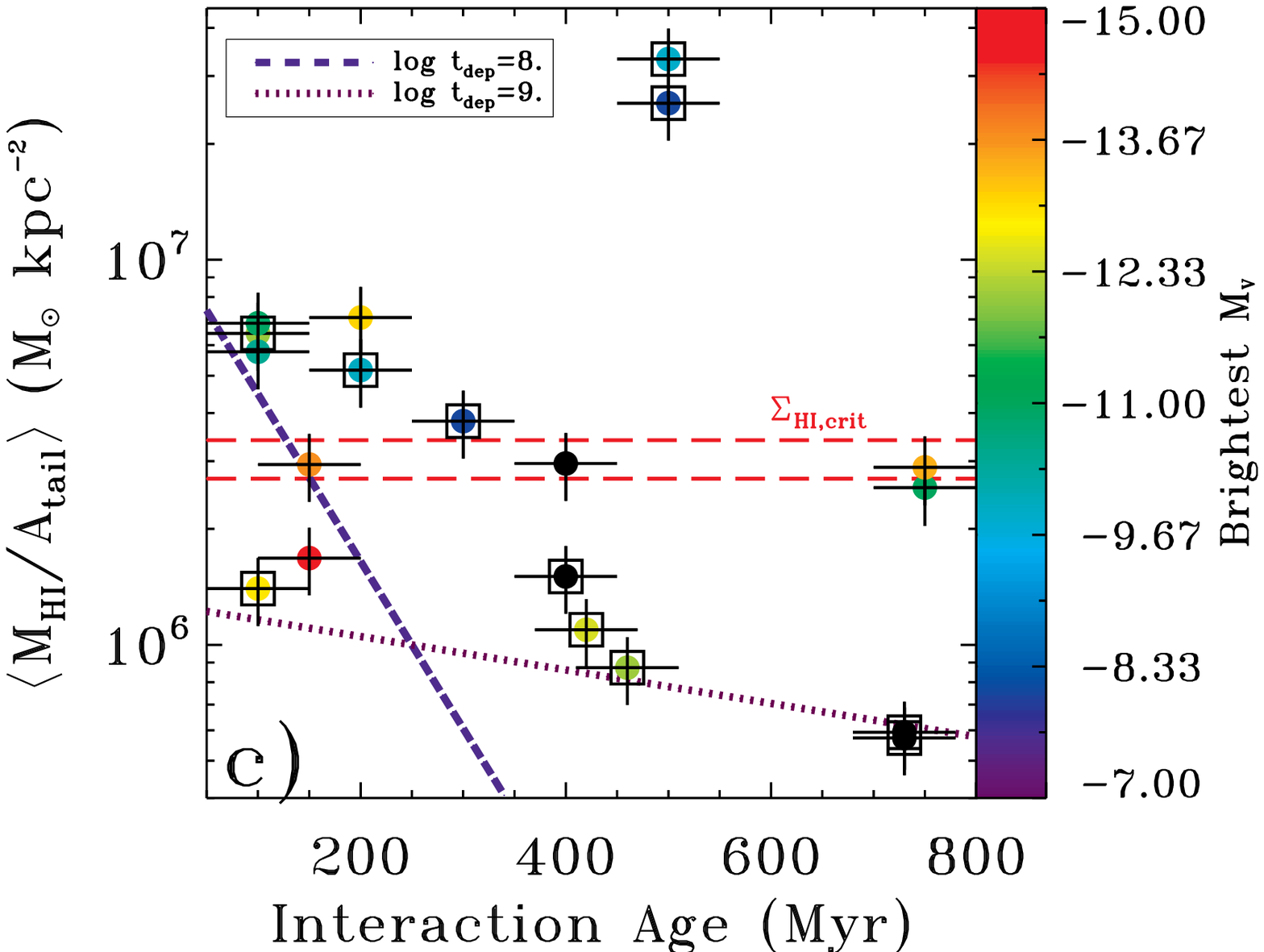,width=0.45\linewidth,clip=0} &
\epsfig{file=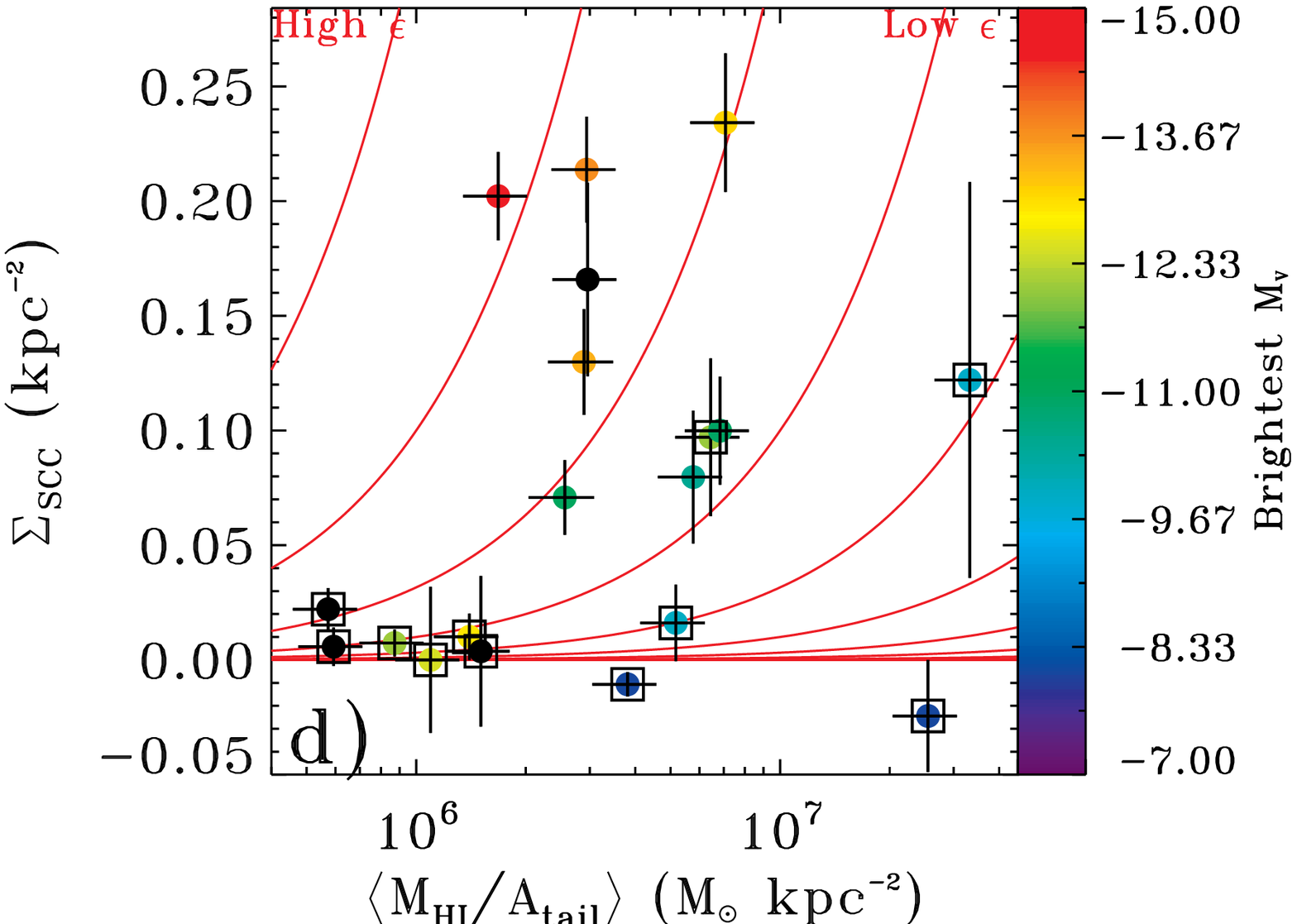,width=0.45\linewidth,clip=0} \\

\end{tabular}
\caption{a) \SCC\ vs.\ \mhi, with colored points corresponding to \HIdens. b) Tail CMD, with points color-coded to \HIdens. c) The distribution of \HIdens with interaction age, with points showing the M$_V$ of the brightest SCC. Dashed lines highlight the range of \HI\ critical densities calculated with the prescription of \citet{schaye}, and dotted lines show hypothetical gas depletion timescales t$_{dep}$ for reference. See text for details. d) \SCC\ vs.\ \HIdens\ with colored points corresponding to the brightest cluster M$_{V}$ detected. Lines trace simple isoefficiency curves of SCC formation with \HI\ density (\SCC\ $\sim$ \HIdens$^N$ for $N$=1). Boxed points in all plots indicate sources with SCC densities statistically equivalent to 0.0 and all error bars show 1$\sigma$ uncertainties.}
\end{figure*}
 \addtocounter{figcount}{1}

\mbox{Figure 7a} first offers a tail CMD whose points correspond to their \HIdens\ values. As in \mbox{Figure 6a}, tails with and without clusters span a range ages and \HIdens\ with no obvious relationship to integrated $V$ or $V$-$I$.  Examining the interplay of \HIdens, \mhi, and \SCC\ reveals a different picture, however. \mbox{Figure 7b} shows the variation of \SCC\ with \mhi, with data points colored according to a scale of \HIdens. Despite an overall increase of \SCC\ with tail \HI\ richness, there is too much scatter for \mhi\ to be a useful diagnostic of cluster candidate excess. But between the boxed and unboxed data sets, tails lacking SCCs (boxed) appear generally less \HI-dense on WFPC2 FOV scales than SCC-containing tails (unboxed). Without the aberrant NGC 1487E/W tails, K--S tests compute that tails without SCCs have a probability of 0.0216 of being drawn from the same \HIdens\ distribution as those with SCCs. 

Thus, while \HIdens\ is \textit{not} a direct measure of the sub-kpc scales that are important in star formation (e.g.\ \citealp{aparna07}; \citealp{bigiel08}; \citealp{schaye}), and the differences in \HIdens\ populations are borderline distinct, it may be a useful zeroth order diagnostic. Cluster-deficient tails overall appear to have a different distribution of \HI\ surface densities than tails with SCC excesses. Many of the boxed points in \mbox{Figure 7a} are bluer -- have lower \HIdens\ -- than the normal points, although there is scatter (e.g.\ the \HI-dense and highly inclined NGC 2992 whose SCCs are largely confined to its TDG).
       
It is therefore interesting to consider whether there is some cutoff value or transition range of \HIdens\ that would account for the different distributions. According to \citet{schaye}, it is possible to estimate a critical gas surface density for star formation, whose \HI\ component we would observe here. Varying the input parameters of the \citet{schaye} models (see their Equation 25) and estimating the contribution of molecular gas yields plausible critical \HI\ surface densities $\Sigma_{\mbox{\scriptsize{\HI\rm{, crit}}}}$ $\approx$ 2.7--3.4 $\times$ 10$^6$ \msun\ kpc$^{-2}$. Here we have assumed mean metallicities $\lesssim$ Z$_{\odot}$, and \HI-ionizing UV radiation fields of 10$^4$--10$^6$ photons cm$^{-2}$ s$^{-1}$, which are reasonable estimates for tidal tails given their range of luminosities, ages, and sizes. We estimate that H$_2$ contributes no more than 10\% to the total gas density from \mbox{Figure 13} of \citet{bigiel08}. We will hereafter refer to 2.7--3.4 $\times$ 10$^6$ \msun\ kpc$^{-2}$ (2.7--3.4 \msun\ pc$^{-2}$) as the fiducial critical gas density for star formation in tidal tails. Again, it is crucial to stress that a threshold density technically applies to sub-kpc scales, not the mean \HIdens\ values quoted here. However, tails like NGC 2444, NGC 7252E/W, and NGC 3921S, with \HIdens\ $\lesssim$ 1.7 $\times$ 10$^6$  \msun\ kpc$^{-2}$,  \textit{on average} may be unable to generate significant SCC excesses on WFPC2 FOV scales. This does not prohibit more localized concentrations of cluster-forming activity, but may just imply an overall suppression of cluster formation.   

To further visualize the role of WFPC2 FOV-scale \HI\ density, \mbox{Figure 7c} relates \HIdens\ to interaction age. \HIdens\ declines with respect to the interaction age (again, the anomalous NGC 1487E/W debris should be overlooked). Across the sample, there appears to be an exponential decline in \HIdens\ with time; we plot example curves of \HIdens\ $\propto$  e$^{-t/t_{\rm{dep}}}$ for two depletion timescales $t_{\rm{dep}}$. We include t$_{\rm{dep}}$ = 100 Myr (blue dashed lines) and 1 Gyr (violet dotted lines). There are necessarily strong inter-tail differences in depletion timescales from star formation, orientations, and time-dependent gas dispersal. There is also a distance-related \HIdens\ bias -- clumps of \HI\ emission at different places along a tail can influence \HIdens\ measurements depending on the extent of coverage and placement of the WFPC2 FOV, and therefore distance. But in general, an average depletion timescale of $\sim$ 500 Myr for these tidal tails may be reasonable. This is on par with typical tidal tail lifetimes \citep{binney08}.

In addition, red dashed lines in \mbox{Figure 7c} highlight our estimate of $\Sigma_{\mbox{\scriptsize{\HI\rm{, crit}}}}$. Evidently the notion of a critical \HI\ surface density does not successfully translate to a critical \HIdens\ for cluster formation. There are several tails (e.g.\ NGC 6872E/W) around or below this \HI\ limit with an ample supply of clusters. Conversely, several tails above the $\Sigma_{\mbox{\scriptsize{\HI\rm{, crit}}}}$ limits are statistically lacking in clusters. Each point in \mbox{Figure 7c} glosses over the particular dynamics of each interaction that obtain varying quantities of the gas, the true three-dimensional densities and pressures of the tail ISM, and the multiple generations of stars and clusters that form a complex recycling and feedback network with said ISM. Thus there are clearly more complicated and local physics at work in determining cluster formation thresholds.

The points plotted in \mbox{Figure 7c} are colored according to the corresponding scale of the brightest cluster candidate M$_V$. Between the tails with and without significant cluster excesses, we find a K--S probability of 0.0224 for their brightest M$_V$ values being drawn from the same distribution. Like the K--S results for \HIdens, the differences in the brightest M$_V$ distributions are marginal, but probably real. From the size-of-sample effect (e.g.\ \citealp{larsen04}; \S3.3), tails with cluster candidates likely have strong local star and cluster formation rates, and therefore have more luminous brightest cluster magnitudes. This does not seem to conspicuously relate to \HIdens\ or interaction age, again obfuscating any additional attempts to connect overall tail star and cluster formation with dynamical age and large-scale \HI\ content. 

Lastly, \mbox{Figure 7d} compares \HIdens\ to \SCC, using the colors of the plotted points to again represent the brightest cluster candidate magnitudes for each tail. Since most of these tails are presumably produced from the \HI-dominated gas of their progenitor galaxies' outer regions, \HIdens\ may approximately represent the total mean gas density. If tidal tails behave like spiral galaxies, the area-normalized tail star formation rate $\Sigma_{SFR}$ may possibly then scale with \HIdens\ by a Schmidt-Kennicutt law; $\Sigma_{\rm{SFR}} \propto$ \HIdens$^{\rm N}$ for some index N \citep{KSlaw}. If we assume from the size-of-sample effect that the total number of clusters produced a tail scales roughly with the tail SFR, then \SCC\ $\propto$ \HIdens$^{\rm N}$. N=1 implies a constant cluster formation efficiency; curves of constant cluster formation efficiency $\epsilon$ are indicated on the plot. Steeper curves for a certain average \HI\ surface density indicate higher cluster formation efficiency. We avoid reporting efficiencies directly, as they require a conversion of \SCC\ to the total cluster mass density with a currently unreliable estimate of cluster population masses. In reality, the translation from \SCC\ to $\Sigma_{\rm{SFR}}$ would include a treatment of how the luminosity evolution of clusters from earlier episodes and their appropriate timescales affects the measured SCC count, but this is an acceptable approximation given our other uncertainties.

In comparing sub-kpc \HI\ surface density to SFR density for a set of 18 nearby normal and \HI-dominated galaxies and multiple resolution scales within them, \citet{bigiel08} demonstrate that neither the total gas density nor the \HI\ surface density is the critical quantity in determining the SFR. Many galaxies can span about an order of magnitude in star formation efficiency for constant values of \HI\ surface density. Employing similar methods to study regions beyond R$_{25}$ in galaxy disks, however, \citet{bigiel10} regain a Kennicutt-Schmidt scaling of the SFR to \HI\ density, reporting a suppression of star formation efficiency relative to the galaxy interiors. Our data do not show such a scaling, but rather tails spread across different cluster formation efficiencies and \SCC\ values for limited ranges of \HIdens. The Schmidt-Kennicutt law applies strictly to rotating disks, and not tidal tails, so it is either possible no similar relationship would be observed between \SCC\ and any variety of gas for these environments at any scale, or a more directed, local study akin to \citet{bigiel10} and \citet{aparna07} would be more illuminating.

For now, it is interesting to note that tails with the most luminous brightest cluster magnitudes tend to lie around curves of higher $\epsilon$, implying that some of our more cluster-rich tails (NGC 6872E/W and NGC 1614N/S) have generally higher star formation efficiencies (with respect to measured \HI\ reservoirs) than other tails. But ultimately, the scatter in all the plots of \mbox{Figure 7} implies that cluster formation in tidal tails is determined less by \HI\ richness and mean density on WFPC2 FOV scales, and more by other variables that set star formation efficiencies and gas consumption timescales. While cluster-containing tails tend to have higher observed \HI\ densities (often combined with young dynamical ages and bright, star-forming environments) than those without them, the number of cluster candidates observed in these tails cannot be well predicted by our current \HI\ properties. Complex gas dynamics, star formation histories, and tail orientations do not allow a determination of critical gas properties required to produce SCCs. In an upcoming work, we will concentrate on \HI\ densities on more localized scales, and may better uncover the role of neutral hydrogen in cluster formation within tidal tails.

\subsection{Cluster Candidates at Fainter Magnitudes}

Of the 23 tails examined, 10 have nonzero SCC excesses at M$_V <$ -8.5, the range of which has few obvious dependences on many observable parameters. Provided the implications of \mbox{Figure 4} and \mbox{Figure 5}, fainter clusters are more numerous, and in the closest tails, detectable. While we cannot infer LFs and MFs from the current data, it is interesting to ascertain if star clusters exist to the faintest possible limit in each tail. In short, do star clusters of some magnitude form in all tidal tails?

Ideally, to detect the maximum number of cluster candidates in each tail, changing the magnitude criterion to the completeness limits of the tails is desired. The generally higher \SCC\ values for M$_V <$ -6.5 in \mbox{Figure 4} are illustrative of this concept. On the other hand, extending M$_V$ to fainter limits risks contaminating source lists with bright, high-mass main sequence stars from current bursts of star formation, as well as post main sequence stars above a certain mass around the red supergiant stage (e.g.\ \citealp{efremov87}). The combination of these two effects is a strongly age-dependent function of the tails' star formation histories and the minimum single stellar mass that the new M$_V$ or completeness limit provides. The blue and red (M$_V <$ -7.5 and -6.5) points in \mbox{Figure 4} cannot be interpreted without this caveat.

Appendix B explores these topics in further detail for the interested reader. Combining the extended magnitude limits with a treatment of single stellar contamination, we find that the NGC 520 TDG candidate probably has clusters to M$_V$ = -6.5, when there were no clusters brighter than M$_V$ = -8.5. NGC 1487E/W and NGC 4747 are also complete to fainter magnitudes, but their statistics are too poor for interpretation or their potential contamination was not negligible. For more distant tails, it cannot be determined for certain whether they have cluster excesses at fainter magnitudes than M$_V$ = -8.5.  Some tails have insightful clues to their dearth of cluster candidates, however. We offer an explanation of some of these tails below.

\textbf{NGC 1487E/W, NGC 4038, NGC 3256E, NGC 7252E/W, and NGC 3921:} It may be no coincidence that these mergers, which have no significant SCC excess at M$_V <$ -8.5, are all $>$ 400 Myr old. A current low tail SFR and significant fading for any clusters formed near periapse may combine to generate an apparent paucity of star clusters. It is not necessarily true that there are no clusters that were formed in the these tails, but rather that the few that are there do not create a statistically significant excess above the background level at the magnitude limit of the WFPC2 images. Their cluster formation events may have been too weak, poorly timed, and/or poorly placed along the tails (the K03 tail pointing is a small fraction of the full length of the NGC 4038 debris region; see their \mbox{Figure 1}). \citet{renaud09} provide a useful illustration of this age effect with their model of compressive and extensive tides in NGC 4038/9 (their \mbox{Figure 3}). If some fraction of cluster formation occurs where compressive tides arise in tidal tails, the nature of the tidal field may be a useful tracer of the star formation history. Their simulations show that, while the galaxy interiors always enjoy compressive tides along with the observationally obvious signatures of continuous nuclear star formation, these tides in the tails are sporadic at best. They preferentially occur at early ages ($\lesssim$ 100 Myr), and will only occur at later times at particular points along the tails.

\textbf{NGC 2444:} This system, whose tail definition was assigned based on its \HI\ debris and had no detectable optical debris in WFPC2 images, may similarly be devoid of significant star formation. This tail has the lowest recorded \HIdens, 1.34 $\times$ 10$^6$ M$_{\odot}$ kpc$^{-2}$, of the tails in the sample; this system's \HI-dependent tail definition warrants caution in comparing its mean \HI\ density to those of the rest of the sample, however. Even if this medium were gravitationally bound, the average free fall timescale (1/$\sqrt{G\rho}$) would be $\gtrsim$ 1 Gyr, longer than the estimated tail age (100 Myr). So overall, star formation may be suppressed. Localized concentrations of \HI\ on sub-WFPC2 FOV scales (kpc$^2$, as opposed to the current $\approx$ 400 kpc$^2$) may reveal more reasonable collapse timescales, however. 

Numerical models of \citet{duc08} suggest that high-velocity encounters (V $\sim$ 1000 km s$^{-1}$) between galaxies generally produce asymmetric, low-mass, gas-dominated tails without much star formation, as opposed to the optically bright (and star-forming), clumpy tails generated by slower mergers. Their model of NGC 4254 reproduces the 250 kpc-long \HI\ stream with a rapid encounter velocity and mass ratio identical to that of NGC 2444/5. Their prototypes of these kinds of collisions produce average \HI\ surface densities $\sim$ 2 \msun\ pc$^{-2}$ or 2 \msun\ $\times$ 10$^{6}$ kpc$^{-2}$, comparable to our result for that quantity.

\citet{app} posit that the diffuse \HI\ plume of NGC 2444 is a result of a recent ($\sim$ 100 Myr), low impact parameter collision between NGC 2444 and a low surface brightness, \HI-rich companion (not NGC 2445, which still interacts with the system). That unseen companion has since disrupted and its stellar component dispersed enough to fall below our optical tail detection threshold; they cite the velocity structure of the bloated bubble structure to the north of our WFPC2 FOV (online \mbox{Figure 4.12}) as kinematic evidence of this event. This model agrees with the timescales at play in this tail. To produce a $\sim$ 100 kpc long \HI\ streamer in 100 Myr (the time required for an expanding \HI\ shell in the tail to reach its current size from its velocity maps), collision speeds $\gtrsim$ 1000 km s$^{-1}$ are needed, rather than the tidal removal of disk gas comparable to orbital speeds at the outer disk radii.  

This suggests that merger dynamics play a dominant role in determining the SCC-harboring capacity of tidal tails. High-speed galaxy encounters may produce low-density tidal gas whose conditions, i.e.\ low metallicities, H$_2$ deficiencies, and high velocity dispersions, are unfavorable for star formation in detectable quantities.

\textbf{NGC 2782W:} It is interesting that the eastern tail of NGC 2782 has a profusion of in-tail sources ($\approx$ 0.2 kpc$^{-2}$), while the western \HI-rich, optically faint tail does not ($\approx$ 0 kpc$^{-2}$). Optical colors and \HIdens\ measurements are similar for these tails ($V$-$I \approx$ 1 and \HIdens\ $\approx$ 5--8 $\times$ 10$^6$ \msun\  kpc$^{-2}$). The main difference, aside from tail length, is optical brightness. The eastern tail is $\approx$ 1.5 magnitudes arcsec$^{-2}$ brighter in $V$ than the western tail. The physical reason for the difference between the two tails has been explored. \citet{smith94} predicts that the eastern tail is predominantly the consumed, optically bright merger remnant of NGC 2782's companion, while the western tail is the metal-poor debris left behind from the outer regions NGC 2782 itself. Therefore, the eastern tail has an ample molecular hydrogen supply from a merging galaxy interior to fuel its observed SCC density. 

The absence of CO emission and inferred lack of H$_2$ in the western tail makes the apparent absence of young cluster candidates understandable \citep{braine01}. A low metallicity of this tail, and differences in the survivability of H$_2$ and CO at low tail extinctions, may account for a nondetection of CO and an inaccurate translation to molecular hydrogen. But if the absence of cluster candidates in this tail is a question of gas availability, projection effects may be to blame. Theoretical models require large relative inclinations between the two galaxies \citep{smith94}, resulting in an extended \HI\ plume with a sizable component along our line of sight. Thus, this system's gas surface density may represent the same physical density as a ``flatter" tail with a lower surface density by as much as a factor of $\sim$ 3. 

Using Gemini optical spectroscopy, Torres-Flores et al.\ (2011, in preparation) confirm cluster status for several objects in NGC 2782W we label as SCCs, and assign them masses $\sim$ 10$^5$ \msun\ and ages consistent with \textit{in situ} formation. Since there are still only a few clusters in the $\sim$ 10$^5$ \msun\ range, so this system suffers from a ``weak" cluster formation history compared to many of the other tails studied here, perhaps for reasons proposed above. NGC 2782W therefore has clusters, but they are effectively hidden in the statistics of our study.  

In summary, certain merger dynamics like encounter velocity, or variations in molecular gas content may account for some of the dispersion in \SCC\ and the range of cluster magnitudes possible across tidal debris. This could be especially pertinent for tails that have no demonstrable SCC excess even at the faintest M$_V$ limits. Variations in completeness (distance) and limited prior observations and theoretical work make this difficult to assess across the entire tail sample. Finding SCCs may be optimal in the closest, youngest tails that are actively star-forming. Disentangling other, potentially more promising effects provided by dark matter halos, merger geometries, etc.\ require further efforts in high resolution modeling.



\section{Conclusions}

We have presented a survey of the star cluster candidates in 17 tidal tails of 12 interacting galaxies observed with {\it HST}/WFPC2 in \Vhst\ and \Ihst. We combined this sample with the six tails studied by \citet{K03}. The 23 tails span the vast parameter space of interaction characteristics like ages, mass ratios, average \HI\ surface densities, \HI\ richness, tail lengths and surface brightnesses, and host galaxy star formation rates.

Star cluster candidate (SCC) densities in the tail regions were computed by subtracting a background estimated from out-of-tail sources. A significant excess ($>$ 2.5$\sigma$) was detected at M$_V <-8.5$ and $V$-$I$ $<$ 2.0 for 10 of 23 tails. This color and magnitude range was selected to ensure completeness, exclude single stellar contaminants, and allow for stochastic and systematic effects of cluster evolution. We have excluded tails whose SCCs were largely constrained to tidal dwarf galaxy (TDG) candidates. In some cases cluster formation seems confined to these structures, as was seen in \citet{K03}. We also identify several cases with definite in-tail SCC excesses and TDG candidates, but cannot extrapolate the significance of this with the paucity of multiwavelength and kinematic data for several of these systems. We also find three cases of ``beads on a string" morphology (with separations of $\approx$ 3 kpc) for clumps of sources/SCCs, characteristic of large-scale star formation by gravitational instabilities also seen in spiral galaxies and tidal tails of other studies.    

For some nearer tails, fainter sources could be detected, and in the NGC 520 TDG candidate a significant cluster excess was found for M$_V <-7.5$ and M$_V<-6.5$, where none was detected at M$_V < -8.5$. Given the limitations of our exposure times and broadband coverage for this project, it cannot be conclusively determined if the in-tail cluster candidate excesses at fainter magnitudes are significant for other tails, or if more distant tails have faint clusters that are currently undetectable. 

To the extent that we can determine them with small numbers, the cluster luminosity function in the tail environment may follow the same power-law slope as in quiescent and starburst galaxies. Combined with the emergent consensus on the universality of clustered star formation, we therefore contend that cluster populations are likely to exist to some level in most, if not all tails. This level may not always be detectable by \textit{HST}.  

A complex combination of various factors, including environmental differences in star formation, gas supply, tail ages, and projection effects appear to influence the populations we observe. Of the global properties we considered (interaction age, progenitor mass ratios and star formation rates, and total \HI\ tail masses), no parameter was individually responsible in shaping the cluster populations. However, we do find statistical differences in the populations or tails with and without cluster excesses in terms of $V$-band surface brightness and mean \HI\ density. The tails that have the highest excesses of star cluster candidates at M$_V <$ -8.5 tend to be the youngest ($\lesssim$ 250 Myr), probably tracing the strongest episodes of star formation that are typically triggered close to periapse. Young tails also tend to have larger mean observed \HI\ surface densities than older tails. It is important to note, however, that there are exceptions to all of these trends.  

We conclude that there is no single explanation for a tail to have a cluster excess, though there are several important influential factors like local gas density and local star formation activity at the time of observation. Quantities like magnitude of the brightest observed SCC and the overall tail surface brightness appear to loosely trace SCC excess. Merger dynamics not addressed here (e.g.\ encounter speeds and orientations; dark matter halo structure) may also be important in imprinting the resulting tidal field with a dynamical setup conducive to star and cluster formation. To disentangle the full assortment of interaction variables would be aided by a larger, deeper sample, more accurate cluster age-dating, and by a higher resolution, local study of the relationship between gas and star cluster formation.

\acknowledgments
We wish to thank B\"arbel Koribalski for sharing \HI\ data and helpful comments for NGC 6872. This projected was supported by a grant from the Space Telescope Science Institute (grant no.\ HST-GO-11134.05-A). 

{\it Facilities:} \facility{HST (WFPC2)}, \facility{VLA}, \facility{ATCA}


\appendix



\section{A. Notes on Individual Systems}

Below we present observations of each tail. Note that interaction ages without citations were calculated by dividing the projected length of the tail by an estimate of the escape or rotation velocity. All observed and measured characteristics quoted are recorded in Tables 2--4. In the WFPC2 images and color maps, the second, third, and fourth chips (WF2, WF3, and WF4), are located in the lower left, lower right, and upper right parts of the mosaics when positioned as in Figure 3.1. 

\subsection{NGC 1487 (Figures 3.1--3.2)} 

NGC 1487 is the optically bright remnant of a $\sim$ 500 Myr-old merger between two comparably massed galaxies \citep{lee05}, whose eastern tail features a long ``spine" curving into the bottom right corner of WF3, surrounded by an extended plume that reaches into the other chips. There is one non-tail source and zero in-tail sources meeting the cluster selection M$_V$ and $V$-$I$ criteria. The in-tail color distribution of detected objects is very broad, peaking at \VI\ $\sim$ 0.5--0.8. The bluest sources of this distribution are concentrated in the blue knots evident in this tail's color map, which trace the spine to the collection of sources at the tail's tip. These may be either very young ($\lesssim$ 10 Myr) or metal-poor.  

There is considerable scatter in the concentration index plot.  At this distance, the largest sources may be marginally resolved (1 pixel = 5.22 pc), so these concentration indices may loosely correspond to a range of physical sizes. For most in-tail sources, the $V$-$I$ colors indicate ages $\sim$ 100 Myr, so many faint sources detected in this study may be clusters $\sim$ 10$^4$ \msun. Thus, we see evidence of star formation around the time of perigalacticon, with more recent, superimposed populations across the tail. These regions in particular paint a high-resolution ``beads on a string" across the color map. Considering possible projection effects, these young clumps are separated by $\sim$ 2--3 kpc.

K--S tests reveal a 34.1\% and 2.63\% probability that all the in-tail and out-of-tail sources detected originate from the same distribution of \Vhst\ and \VI, respectively. The lack of SCCs is strange, given the high average \HI\ densities measured for this tail (\HIdens\ = 2.55 $\pm$ 0.51 $\times$ 10$^7$ \msun\ kpc$^{-2}$). Overall, the tail has one of the lowest total \HI\ masses, however; \mhi\ = 6.14 $\pm$ 1.2 $\times$ 10$^8$ \msun. Why the debris with the highest \HI\ content should also be devoid of cluster candidates $\gtrsim$ 10$^5$ M$_{\odot}$ is not easily explained.  In \msun\ pc$^{-2}$, \HIdens\ = 25.5 $\pm$ 5.1, well above the \HI\ saturation threshold of 9 \msun\ pc$^{-2}$ of \citet{bigiel08}. Above this limit, most of the gas available in the tail may be molecular, so inclination effects or the superposition of disk and tail gas may be responsible in artificially increasing the observed  average \HI\ density.

If the true \HI\ surface density of NGC 1487E is significantly lower and in the expected \HI-dominated regime of typically low-Z, H$_2$ poor outskirts of galactic disks \citep{bigiel08}, the overall SFR of the tail may be relatively low (c.f.\ the host galaxy \textit{IRAS} SFR of 0.12 \msun\ yr$^{-1}$). This may imply an underpopulation of the high-mass end of a star-forming generation's cluster initial mass function (CIMF). This would then result in zero objects observed with photometric characteristics of $\sim$ 10$^5$ \msun\ clusters.

It is easy to check on the logical consistency behind this reasoning. If we reasonably assume that these clusters are drawn from a power-law luminosity function with a slope of -2 -- -2.5 (\S 3.3), then it is possible to analytically estimate how many clusters we should observe brighter than a certain magnitude. Using the number of sources we observe between adjacent, complete luminosity bins (of size 1 mag) to normalize the LF, we calculate that there should be $\lesssim$ 1--3 sources brighter than M$_V$ = -7.5, and no objects brighter than M$_V$ = -8.5. With the tail age limits, this verifies that there should not be clusters present with masses $\gtrsim$ 10$^4$ \msun\ in this tail. 

The western, also \HI-dense yet \HI-poor (\HIdens = 33.2 $\pm$ 6.6  $\times$ 10$^6$ \msun\ kpc$^{-2}$; \mhi = 5.41 $\pm$ 1.1 $\times$ 10$^8$ \msun), tail is a patchy assembly of diffuse light stretching across WF3 and WF4. Similar star-formation hotspots as in the eastern tail are observed, but are not identified as SCCs. There are two cluster candidates found within the tail, and zero beyond it. One SCC has a magnitude and color consistent with a $\sim$ 10$^5$ M$_{\odot}$ cluster from the same $\sim$ 100--300 Myr generation as most of the sources fainter than the M$_V$ selection limit. The color map shows most of the cluster and field light is produced from this generation of star formation, with scattered bursts of younger and less massive ($\lesssim$ 100 Myr; $\sim$ 10$^4$ \msun\ ) objects. This is qualitatively similar to the eastern tail.

There is a 59.3\% and 39.6\% probability of in-tail and out-of-tail sources being drawn from the same \Vhst\ and \VI\ distribution. It is possible that the true tail boundary extends to the \HI\ contours pictured in the online Figure, with several out-of-tail sources belonging to the in-tail source list. However, if this were true, the out-of-tail CMD still indicates that there would still be only two SCCs counted in this tail. The concentration index plot shows no useful distinction between in-tail and out-of-tail detections.

In the galaxy interior, \citet{Mengel} investigate several clusters whose discrepancies between photometric and dynamical masses offer insight into their current disruption. Their clusters are $\approx$ 8.5 Myr old and between 10$^5$ and 10$^6$ \msun. \citet{lee05} performed WFPC2 $B$- and $I$- band photometry on 560 clusters similar to this study, and while their age and mass estimates for clusters are also limited to two bandpasses, they suspect a similar, roughly bimodal age distribution in the NGC 1487 clusters, peaked at 15 Myr and 500 Myr.

\subsection{NGC 4747 (Figure 3.3)} 

This system is a $\sim$ 320 Myr old minor merger interacting with NGC 4725 (\citealp{haynes79}; \citealp{wevers}), featuring a long, optically faint tail extending through the third and fourth WFPC2 CCD chips. There is a concentration of sources in the third chip, with estimated ages $\lesssim$ 50 Myr (and corresponding masses $\lesssim$ 10$^5$ M$_{\odot}$). Of the sources shown in the CMDs, 7 and 5 cluster candidates were detected in and outside the tidal regions, respectively, with \VI\ $\sim$ 0--0.7 and 0.3--0.7. Population differences may be seen in the $\Delta_V$ diagrams; in-tail sources show a slight preference for $\Delta_V$ $\lesssim$ 2.0, compared to the more even dispersion for objects beyond the tail. Given the distribution of colors, SSP models posit ages of $\sim$ 30--300 Myr for $\sim$ 10$^5$ M$_{\odot}$ cluster candidates in this image.

The \Vhst\ and \VI\ K--S probabilities for all sources are 0.878 and 0.646, respectively. \HI\ was not detected in this tail, and the interior of the merger has an \textit{IRAS}-derived SFR of $\approx$ 0.27 \msun\ yr$^{-1}$.

 \subsection{NGC 520 (Figure 3.4)} 

NGC 520 appears to be a minor merger remnant \citep{stanford91}. WFPC2 images present an optically faint, \HI-defined (\mhi\ = 1.12 $\pm$ 0.22 10$^9$ \msun) curved tail extending to the north. The \textit{HST} pointing here is situated on the tidal dwarf candidate in this system (UGC 957), which may either be a genuine TDG or a small satellite galaxy engaged in a 3-body interaction with the other galaxies \citep{stanford91}.  This object coincident with bright \HI\ emission (\HIdens\ = 3.81 $\pm$ 0.76 $\times$ 10$^6$ \msun\ kpc$^{-2}$), but no CO has been detected at this location \citep{stanford91}. 

Within the host galaxy, infrared emission indicates an SFR of $\approx$ 6.5 \msun\ yr$^{-1}$. Given the TDG candidate's distance from the galactic interior (112 kpc), the tail/galaxy SFRs may not correlate well. GALEX observations \citep{Neff} have revealed that no current star formation has occurred within NGC 520, and the youngest regions are $\sim$ 300 Myr old. Moving out to the tail, they find massive 10$^8$--10$^9$ M$_{\odot}$ clumps of ages $\gtrsim$ 300 Myr. Optical spectroscopy of the NGC 520 nucleus confirm a 300 Myr old episode of star formation, with little current activity \citet{stanford91a}. This matches dynamical simulations of this interaction \citep{stanford91} that place the age of this interaction at 300 Myr. 

The distribution of colors detected in the TDG is extremely broad in the CMD and color map and lacks an obvious peak. We detect 0 and 4 SCCs in and beyond the tail, respectively. Given the above age constraint, many sources fainter than M$_V$ = -8.5 may be $\sim$ 10$^4$ M$_{\odot}$ clusters $\lesssim$ 10 Myr old, or $\lesssim$ 5 $\times$ 10$^4$ \msun\ clusters with ages $\sim$ 150--500 Myr. 

From the $\Delta_V$ diagram in the online Figure, NGC 520 may contain a marginally distinct population of compact objects, whereas the external sources are generally more extended and fainter (M$_V \lesssim$ -7).  K--S tests estimate a probability of 1.37 $\times$ 10$^{-3}$ and 0.757 for all detections originating from the same \Vhst\ and \VI\ distributions.

 \subsection{NGC 2992 (Figure 3.5)} 

NGC 2992 is a peculiar Sa galaxy with an unusual optical condensation at the edge of its tail in WF2. With a combination of H$\alpha$, CO, optical, and \HI\ data, \citet{bournaud04} asserts that this is a genuine TDG. Sources detected within outline prominent CO emission; with standard conversion factors, \citet{bournaud04} estimate that the TDG contains 3.5 $\times$ 10$^7$ \msun\ of H$_{2}$, contributing to a total of $\approx$ 2 $\times$ 10$^9$ \msun\ of stellar and gaseous material.  

There appears to be a deficit of sources in the middle of the tail on WF3, which is visibly smoother and undisturbed. The \VI\ color map is remarkably homogeneous, with colors consistent with ages of $\sim$ 100 Myr -- 1 Gyr. \citet{brinks04} find that this tail has a high metallicity within the TDG (12 + log(O/H) = 8.6), so redder colors are expected. The embedded SCC population with similar \VI\ would be mostly composed of $\sim$ 10$^5$ M$_{\odot}$ clusters. It is also possible from the CMDs that lower mass, $\lesssim$ 10 Myr old clusters contribute to the tail light. Of the sources indicated, 37 and 15 are SCCs in and out of the tail, respectively.  Between these two regions, there do not appear to be distinctions in terms of concentration index. We find K--S \Vhst\ and \VI\ probabilities of 0.0187 and 0.660 for the entire source list. The location of SCCs coincides with the brightest \HI\ contours; this tail has a mean \HI\ density of \HIdens = 6.44 $\pm$ 1.3  $\times$ 10$^6$ \msun\ kpc$^{-2}$. It also has an interior \textit{IRAS}-derived SFR of $\approx$ 3.4 \msun\ yr$^{-1}$.

This galaxy is in the early phases of merging with its similarly-massed companion NGC 2993. They both exhibit moderate \HI\ masses (\mhi\ = 1.43 $\pm$ 0.29 $\times$ 10$^9$ \msun\ for NGC 2992), optical brightnesses, and short (apparent) tails relative to their disks. 100 Myr have elapsed since perigalacticon \citep{duc}, and the final merger should occur in $\sim$ 700 Myr. Given this constraint, SCCs that form in situ must be $\lesssim$ 5 $\times$ 10$^5$ \msun, as mentioned above. Prior studies have noted difficulties in isolating central starbursts in this AGN-powered Seyfert galaxy (e.g. \citealp{davies07}; \citealp{fanelli97}), so the star cluster populations of that system remain unresolved.

 \subsection{NGC 2993 (Figure 3.6)} 

The interacting counterpart to NGC 2992, NGC 2993 features an optically faint, moderately \HI-rich (\mhi = 7.27 $\pm$ 1.5 $\times$ 10$^8$ \msun; \HIdens = 5.77 $\pm$ 1.2 $\times$ 10$^6$ \msun\ kpc$^{-2}$) tail that curves through the second and third chips of the \Vhst-band mosaic. The interior of this galaxy supports an infrared-derived SFR of $\approx$ 5.0 \msun\ yr$^{-1}$. The color map exhibits a strong regional variation of \VI, from the blue spine to the redder outlying plume. 

In the color-magnitude diagram for the debris, in-tail objects are roughly distributed across the tail and its color gradient, and mixed with the out-of-tail sources in the concentration index diagram. There are 15 in-tail SCCs and 7 out-of-tail counterparts, the former of which show a degeneracy of possible high masses/old ages or lower masses and young ages akin to NGC 2992. From H$\alpha$ equivalent widths, \citet{duc} find groups of 5--7 Myr old starbursts in the main body of this system, which are systematically younger than the estimated $\sim$ 10 Myr -- 700 Myr age range of objects detected here. If tail populations of NGC 2992/3 are around the same age, then our detected SCCs may have masses $\sim$ 10$^4$ \msun. K--S tests reveal a 1.72\% and 65.2\% probability that all the in-tail and out-of-tail sources detected originate from the same \Vhst\ and \VI\ distribution, respectively.

 \subsection{NGC 2782 (Figures 3.7--3.8) }

NGC 2782 is an SABa-type, $\sim$ 200 Myr old minor merger \citep{K07}. The eastern tail is short compared to the face-on optical disk, and rich in \HI\ (\mhi\ = 2.41 $\pm$ 0.48 $\times$ 10$^9$ \msun; \HIdens\ = 7.08 $\pm$ 1.4  $\times$ 10$^6$ \msun\ kpc$^{-2}$). The aftermath of the interaction is evident in the disrupted appearance of the main disk in WF4, highlighted by ripples and density inhomogeneities that continue down into the tail in the other two chips. The detected sources seem loosely correlated with the optically clumpy parts of the tidal debris, and appear only partially correlated with the bluer regions in the color map. We find 87 potential clusters in the tail, and 4 beyond.

For all the in-tail objects, the CMD reveals a broad \VI\ distribution of colors representative of a correspondingly broad range of putative cluster ages. SCCs may have a spread of ages and masses, from $\lesssim$ 10 Myr, $\sim$ 10$^4$ \msun\ clusters, to those aged nearly 1 Gyr and 10$^5$ \msun\ in mass. Given the age constraint and 100 Myr old cluster estimates for this tail of \citet{K07}, our SCCs are more likely closer to the former than the latter. 

Contours of \HI\ emission for the eastern tail abruptly end halfway across the third chip; this is also where the bluest (or conversely metal-poor) regions of the tail are. As seen in the online Figure, many of these sources tend to be photometrically extended ($\Delta_V$ $\gtrsim$ 2), similar to the few external sources with generally high concentration indices and more limited spread of colors. Blue colors (\VI\ $\lesssim$ 0) and extended profiles may indicate young ages and concordant nebular emission. The majority of fainter in-tail SCCs appear more concentrated, with $\Delta_V$ $<$ 2.0. We determine K--S probabilities of 0.875 and 0.963 for \Vhst\ and \VI, respectively, for all sources.

The western tail of NGC 2782 is markedly different, barely tracing WF3 to the PC and down into WF2. The optical extent of the tail does not perfectly match the \HI\ tail illustrated on the SDSS image; the latter extends further into WF4 and WF2 than the optical boundary we employ here. This tail is also more \HI-rich on average than its eastern companion (\mhi\ = 3.14 $\pm$ 0.63 $\times$ 10$^9$ \msun; \HIdens\ = 5.17 $\pm$ 1.0  $\times$ 10$^6$ \msun\ kpc$^{-2}$). Its relative paucity of SCCs is in stark contrast to NGC 2782E. With a common progenitor galaxy \textit{IRAS} SFR of $\approx$ 3.7 \msun\ yr$^{-1}$, it is once more apparent that this infrared diagnostic may not correlate with tail SFR in all cases and cannot be interpreted too deeply. 

In terms of concentrations and colors, in-tail and non-tail sources are evenly dispersed about the \VI\ cutoff, and do not show immediately recognizable differences in $\Delta_V$. The ages and masses of these SCCs are similar to those of the eastern tail. Assuming the \citet{K07} inference that the tails are host to population of young clusters of ages $<$ 100 Myr, our sources (if they are clusters) are probably between 10$^4$ and 10$^5$ \msun. 10 SCCs are found in the tail, and 7 are located outside it. Here, K--S tests anticipate a 5.65\% and 2.49\% chance that all in-tail and out-of-tail sources came from a single distribution of \Vhst\ and \VI.

\subsection{MCG-03-13-063 (Figure 3.9)} 

This SBb-type, optically faint galaxy with unknown \HI\ appears slightly perturbed by its $\sim$ 100 Myr old minor merger, with thin filamentary tails drawn from the main disk, enriched with classic ``beads on a string" hotspots of star formation. Assuming an approximately face-on orientation, the separation between these clumps is about 3 kpc. 

The tidal debris and galaxy interior (WF2-WF3) show an extended distribution in \VI, down to very blue (\VI\ $<$ -0.25) colors indicative of young ages and nebular emission not tracked by the SSP models. These sources overall are uniformly distributed in the $\Delta_V$ diagram. Since any clusters would be unresolved at this distance, some may be crowded cluster complexes rather than individual clusters. CMDs show these objects as potentially very young (several $<$ 10 Myr and others $\lesssim$ 100 Myr), and massive. With in situ age constraints, these SCCs would be $\sim$ 5 $\times$ 10$^4$--10$^5$ \msun\ . 

In total, 88 cluster candidates are found in the tails, and 26 are beyond them. The probability that all in-tail and out-of-tail sources originate from one population is 0.324 for \Vhst\ and 0.0617 for \VI. In spite of this overwhelming evidence for clustered star formation in this tail, the \textit{IRAS}-derived SFR is one of the lowest of the tail sample, at $\approx$ 0.43 \msun\ yr$^{-1}$. This system is aligned nearly face-on, and may register lower IR SFRs than galaxies with higher inclinations.

\subsection{ESO 376- G 028 (Figure 3.10) and AM 1054-325 (Figure 3.11)} 

AM 1054-325 is an optically bright spiral in the early stages  (age $\approx$ 85 Myr) of merging with a nearby early type galaxy ESO 376- G 028, as displayed in the online Figure. The latter system features four SCCs embedded in a diffuse plume of debris, but the number of equivalent objects per unit area beyond its tidal debris makes it statistically insignificant. The general \VI\ color of this debris is relatively red (\VI\ $>$ 0.8); though \HI\ data do not exist for this galaxy, on morphological and photometric grounds it stands to reason there is simply insufficient gaseous ``fuel" to trigger obvious star formation in what is probably a dry merging system. For ESO 376- G 028, we find \Vhst\ and \VI\  K--S probabilities of 0.445 and 0.291. 

AM 1054-325 is the more remarkable of the two, with large clumps of star-forming debris pulled out from the disk in WF3. \VI\ colors in the tidal debris evidenced in the CMDs and color map highlight young and blue episodes of star formation, and some evince possible nebular emission in their blue (\VI\ $<$ 0) colors. The bluest sources  are closer to the $\sim$ 1--5 Myr estimates of star formation regions within the galaxy by \citet{Weilbacher}, with some leeway in cluster mass.  

The host galaxy IR SFR is $\approx$ 0.64 \msun\ yr$^{-1}$.  The tail appears to have a base color consistent with an age $\lesssim$ 100 Myr. ``Beads" of newer ($\lesssim$ 10 Myr) and/or metal-poor bursts are strewn across the tail. These clumps in the smoothed color map are separated by $\sim$ 2--3 kpc, depending on tail orientation. The $\Delta_V$ diagram shows the strongest disconnect between in-tail and out-of-tail sources thus far, with the former having typically bluer \VI\ colors than the latter. The wide range of concentration indices may again be symptomatic of cluster complexes that are unresolved at this distance. 

180 sources qualify as cluster candidates inside the tail, with 60 outside it. For AM 1054-325 sources, we determine K--S probabilities of 2.36 $\times$ 10$^{-10}$ (\Vhst) and 9.22 $\times$ 10$^{-25}$ (\VI). The above age restrictions allow these SCCs masses $\sim$ 5$\times$10$^4$--5$\times$10$^5$ \msun. The tail's orientation is unknown, however, and may negatively affect the number of SCCs measured. If the debris extends from the center-right of the galaxy disk and curls northward towards the observer, we may witness a superposition of disk-bound and in-tail clusters that combine to give unusually high SCC counts and exaggerate this tail's cluster bearing capacity. Or if what we identify as a tail is partially a spiral arm pulled by the interaction, we would be observing not tidal debris, but galactic cluster candidates shaped by different arm-related triggering processes like density waves.

\subsection{NGC 2444 (Figure 3.12)}

NGC 2444 is a peculiar S0-type galaxy featuring a long \HI\ tidal tail visible in Figure 3.12. \citet{app} suggest that the \HI\ debris is a result from a collision between NGC 2444 and a low surface brightness, \HI-rich companion, and not NGC 2445, which still interacts with the system. That companion has since disrupted and its stellar component dispersed enough to fall below our optical tail detection threshold. This model makes intuitive sense -- collision speeds $\gtrsim$ 1000 km s$^{-1}$ are needed to create a $\sim$ 100 kpc long \HI\ plume in 100 Myr.

Though an \HI\-rich (\mhi\ = 5.20 $\pm$ 1.0 $\times$ 10$^9$ \msun) tail through the WFPC2 field of view is visible in the SDSS image of this tail's online Figure, any optical counterpart is too faint to be contoured by the 1 count + background technique utilized here. As such, we make an exception for this tail and follow the \HI\ extent (average \HIdens\ = 1.40 $\pm$ 0.28  $\times$ 10$^6$ \msun\ kpc$^{-2}$) illustrated in the online Figure to determine the tidal tail boundary.

There are 27 sources meeting the SCC selection criteria within this \HI\ tail, and 15 outside it. In all, they appear to span an age range of $\sim$ 10--500 Myr, depending on masses between 10$^4$--10$^5$ \msun\ . They are also evenly dispersed among the out-of-tail objects in the $\Delta_V$ diagram. K--S tests dictate a 54.5\% and 24.1\% chance that all in-tail and out-of-tail sources came from a single distribution of \Vhst\ and \VI.

In contrast, \citet{app} and \citet{jeske86} find circumnuclear star-forming regions with ages $<$ 30--69 Myr. These are approximately timed with the last passage of the interacting galaxies that is thought to have produced the long \HI\ plume and merger morphology characteristic of collisional ring galaxies \citep{hibyun}. \citet{beir09} investigate the H$_2$ and PAH properties of star-forming knots within the circumnuclear ring, determining ages of 2--7.5 Myr, contrasted with a nucleus age of about 500 Myr.

\subsection{NGC 2535 (Figure 3.13)} 

An \HI-rich (\mhi\ = 4.45 $\pm$ 0.89 $\times$ 10$^9$ \msun), optically bright Sc galaxy $\sim$ 100 Myr past perigalacticon in its minor merger with NGC 2536 \citep{hancock07}, NGC 2535 possesses two main tidal tails; the long northern tail is illustrated in its online Figure. It contains a ``backbone" of clumps from WF3 to WF2, surrounded by more diffuse emission and anchored to an outer spiral arm in WF4. The optical tail coincides with the \HI\ emission seen in Figure 3.13, (\HIdens\ = 6.84 $\pm$ 1.4 $\times$ 10$^6$ \msun\ kpc$^{-2}$). The innermost limit of the tail boundary corresponds to the terminus of the hinge clumps seen at the rightmost side of WF3. These star-forming hotspots accumulate at the low shear tips of spiral arms \citep{hancock09}, marking the end of the galaxy proper. It is a convenient check on the validity of our tail/galaxy distinction. These hinge clumps also appear as pronounced blue condensations in \VI\ the color map.

A number of sources are evident, aside from the SCCs selected. The in-tail objects are very blue overall in the CMD (\VI\ $<$ 0.5), as opposed to the out-of-tail sources that have redder (0.2--2) colors. Both are scattered in terms of $\Delta_V$, however. We determine that the cluster candidates (41 in the tail and 49 outside) can be fit with ages of $<$ 10 Myr for the brightest objects, to $\sim$100 Myr for sources located around the evolutionary tracks. Masses would then range from 10$^4$--10$^5$ \msun\ . Using optical/UV colors and H$\alpha$ data, \citet{hancock07} constrain the ages of star-forming regions within interior features to $\sim$ 9 Myr for the $\sim$ 10$^6$--10$^9$ \msun\ clumps (complexes) they study. They contend that clumps in the tidal regions of this interaction have similar ages, but lower masses (typically by a factor of $\sim$ 10).

The K--S probabilities of NGC 2535 sources are 0.0904 (\Vhst) and 1.16 $\times$ 10$^{-4}$ (\VI). Our galaxy IR SFR calculation estimates a rate of $\approx$ 2.9 \msun\ yr$^{-1}$ in NGC 2535's disk, similar to the 2.0 $\pm$ 0.8 \msun\ yr$^{-1}$ and 4.9 $\pm$ 2.0 \msun\ yr$^{-1}$ SFRs found for the clumps and total system by \citet{hancock07}.

\subsection{NGC 6872 (Figures 3.14--3.15)}

NGC 6872 is a large, optically bright SABc galaxy well endowed with \HI\ in its combined eastern and western tail pointings (\mhi\ = 3.52 $\pm$ 0.50 $\times$ 10$^9$ \msun), though its tremendous size translates this to a low \HIdens\ of $\approx$ 1.5--3 $\times$ 10$^6$ \msun\ kpc$^{-2}$. The galaxy bears long, thin tidal tails from its $\sim$ 150 Myr old interaction with a smaller companion, IC 4970 (\citealp{horellou}; \citealp{horellou07}). The eastern pointing features an intricate stream of tangled hotspots, from irregular clumps to cluster candidates. Sources in all chips have a large spread in \VI\ -- likely from a combination of age and extinction effects -- with several extremely blue objects featuring possible nebular emission. There is a wide variety of concentration indices around \VI\ $\sim$ 0--1, which is not dissimilar to the out-of-tail sources.

Contours of \HI\ on the SDSS image are roughly aligned with the most populated regions in the optical mosaic and color map. The proliferation of SCCs straddling the WF2-WF3 divide are uniformly blue and young or metal-poor. This may be a hotspot of the most recent episode of star formation or evidence of a metallicity gradient across the tail. The latter scenario is observed in the opposite tail by Trancho et al.\ (2011, in preparation); they find a gradual decline in metallicity from supersolar at the tail base to subsolar at the tip. What we witness in the color map of this tail may be the eastern tail's equivalent gradient, portraying a sequence of cluster formation events beginning at the tail base and proceeding towards the tip, using the metal-poor ISM drawn from the galactic outskirts as fuel. By the time star formation has ignited at terminus of the tail, the base may have endured multiple generations of star formation and concordant metal enrichment.    

The western FOV of the tail, dominated by the galaxy body in WF4, has in-tail sources congregating in clumps around the WF2-WF3 divide, a number of objects spread throughout the tidal debris in all chips, and another assemblage east of IC 4970 coincident with a locally dense \HI\ region \citep{horellou07}. Both tail and external sources have a broad distribution of colors similar to that of the eastern FOV. In either case, there are significant numbers of extremely blue objects (\VI\ $\sim$ -0.75--0.0), as well as those with more moderate colors. Tail sources with \VI\ $\sim$ 0.5--1.5 seem more concentrated at $\Delta_V$ = 1.5--2, and the bluest objects tend to have concentration index values $\gtrsim$ 2. Out-of-tail sources are more evenly dispersed in \VI--$\Delta_V$ space.

Source ages in both eastern and western FOVs range from $\lesssim$ 10 Myr for a few of the bluest cluster candidates, to 100--300 Myr for dimmer and/or redder hotspots. The former correspond well with the 10$^4$--10$^6$ M$_{\odot}$,  $<$ 100 Myr old clusters detected and confirmed to be spatially coincident with regions of H$\alpha$ emission (\citealp{bastian05}; \citealp{mihos93}). It is unclear whether these SCCs form a MF with a logarithmic slope of -1.85 $\pm$ 0.11 as the former study suggested with their ground-based, lower resolution work. Many of the sources they pinpoint may be agglomerations of the higher resolution SCCs we define, which may decrease the MF slope to values closer to $\approx$ -2.  

We postulate a mass range of a few 10$^4$--10$^6$ \msun\ for these objects. In the eastern pointing, there are 163 SCCs within the optical tail boundary, and 63 outside of it. NGC 6872W sports 195 SCCs in the tail, and 36 beyond. We find K--S probabilities of 7.26 $\times$ 10$^{-5}$ and 0.209 for all sources in \Vhst\ and \VI for the NGC 6872E FOV. For the western pointing, K--S tests predict a 0.517\% and 11.9\% chance that all in-tail and out-of-tail sources came from a single distribution of \Vhst\ and \VI. Our IR SFR of $\approx$ 2.9 \msun\ yr$^{-1}$ for the central galaxy is significantly lower than U-band estimate of the local tail SFR of $\approx$ 16.5 \msun\ yr$^{-1}$ for the combined debris regions \citep{bastian05}, but the latter value is likely an overestimate (Bastian 2010, private communication). It is likely the ``true SFR" for the tail is closer to those quoted by \citet{mihos93}; $\sim$ 3 \msun\ yr$^{-1}$.

\subsection{NGC 1614 (Figures 3.16--3.17)} 

NGC 1614 is $\approx$ 750 Myr into its merger with a similarly-massed companion (\citealp{neff90} and references therein), a peculiar SBc galaxy with a nearly linear plume of southern debris and a northern disky tail. Both tails are optically bright and contain comparable amounts of \HI\ (\mhi\ = 7.68 $\pm$ 1.6 and 9.64 $\pm$ 1.9 $\times$ 10$^8$ \msun\ for northern and southern tails, respectively). \textit{IRAS}-derived SFRs are problematic for its tidal tails; it is a heavily dust-obscured ULIRG \citep{alonso01} and presents a nuclear cloud-cloud collision triggered SFR \citep{alonso01} of $\approx$ 35. \msun\ yr$^{-1}$. Its \HI\ content is unevenly distributed across the optical debris -- only the northern and southern tips of the northern tail present \HI\ emission (\HIdens\ = 2.56 $\pm$ 0.52  $\times$ 10$^6$ \msun\ kpc$^{-2}$), as does the optically bright ``stub" in the inner southern tail (\HIdens\ = 2.89 $\pm$ 0.59  $\times$ 10$^6$ \msun\ kpc$^{-2}$).

The color map of NGC 1614N displays a diffuse debris whose \VI\ is generally consistent with its 750 Myr age. Individual SCCs are typically bluer (younger), with \VI\ -estimated ages of $\sim$ 10 Myr to 300 Myr. One SCC may be younger than 10 Myr, depending on whether it falls in the $\sim$ 5 $\times$ 10$^4$--10$^5$ \msun\ range of its companions. These objects show no particular $\Delta_V$ preference in the concentration index diagram, but do approximately coincide with the regions of \HI\ emission. In all, there are 33 and 28 SCCs in and outside the tail, respectively. 

The southern tail shows a greater diversity in SCC demographics than the northern debris. Sources populate the in-tail CMD around the 10$^5$--10$^6$ \msun\ SSP tracks, with corresponding ages $\lesssim$ 300 Myr. The bluest and brightest of these are concentrated in the \HI-rich stub forming the base of the tail. Sources outside the \HI\ tail are largely redder, possibly older sources of the aforementioned age range. Like the northern tail, SCCs here do not display any preference for a particular range of $\Delta_V$. We find 38 SCCs in this northern tail.

This system exhibits a $\sim$ 300 pc circumnuclear ring of massive \HII\ regions in the central galaxy \citep{alonso01}, whose photometric properties imply a series of current (5--8 Myr old), massive (5.6 $\times$ 10$^8$ \msun) bursts, over a base population 1 Gyr or older. The bluest sources in the southern tail may be commensurate with this activity; the ages and masses of the remaining SCCs are not well enough constrained to correlate their formation with intragalactic star formation. In the northern tail, \Vhst\ and \VI\ K--S probabilities for all sources are 1.94 $\times$ 10$^{-3}$ and 3.33 $\times$ 10$^{-4}$, respectively. All southern tail sources show \Vhst\ and \VI\ K--S probabilities of 2.61 $\times$ 10$^{-4}$ and 1.27 $\times$ 10$^{-6}$.




\section{B. SCC Detection at Fainter Magnitudes}

\subsection{SCC Detection Biases} 

To understand the presence or absence of cluster candidates at fainter luminosities, several key issues must be addressed -- the rapid color and magnitude evolution of clusters as they age, the overall level of cluster production in the studied environment, and completeness limits of detected cluster candidates. Thus cluster detection is heavily contingent on \textit{age, size-of-sample effects, and distance}. \mbox{Figure B1} more clearly illustrates the combination of these three influences.


\setcounter{figcount}{1}
\renewcommand{\thefigure}{B\arabic{figcount}}

\begin{figure}[htbp]
\plotone{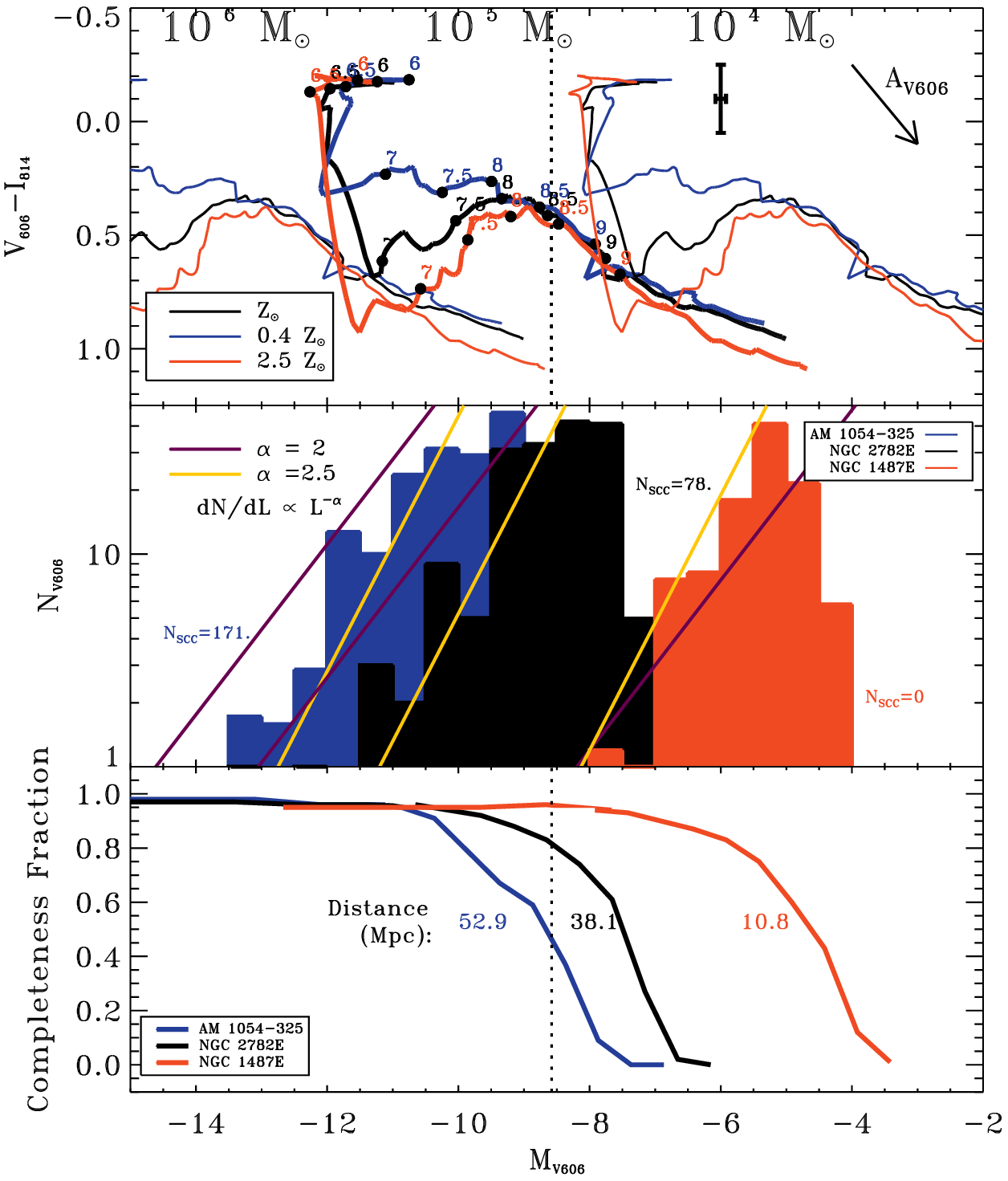}
\caption{\textbf{Top panel:} \VI\ vs.\ \Vhst\ evolution of \citet{BC03} population synthesis models used in this project (see text for details). Different colors represent populations of different metallicities; black: Z$_{\odot}$, blue: 0.4 Z$_{\odot}$, red: 2.5 Z$_{\odot}$. A vector for 1 magnitude of extinction in \Vhst\ and median photometric errors for sources are indicated. Dotted lines mark the color and magnitude selection limits. \textbf{Middle panel:} M$_{V606}$ distribution of statistically background-subtracted objects detected for AM 1054-325, NGC 2782E, and NGC 1487E. The background-subtracted number of SCCs detected is presented, along with artificial luminosity functions for these tails' sources (purple and yellow lines for power-law LFs with slopes -2 and -2.5, respectively). \textbf{Bottom panel:} M$_{V606}$ in-tail completeness curves for the tails of the middle panel. Luminosity distances to host galaxies are shown.}

\end{figure}



The top panel of \mbox{Figure B1} shows the color-magnitude evolution of model clusters from the population synthesis models of \citet{BC03}. The blue track follows a theoretical population with a metallicity 0.4 Z$_{\odot}$, the black shows Z$_{\odot}$, and the red 2.5 Z$_{\odot}$. The tracks are repeated for masses of 10$^6$, 10$^5$, and 10$^4$ \msun; the 10$^5$ \msun\ models are outlined in bold for emphasis. The vertical dotted line in the CMDs indicates where M$_{V}$ = -8.5. For reasonable metallicities, these cutoff values allow the detection of 10$^5$ \msun\ clusters with log age (yr) $\sim$ 6--8.5. Our objective is to observe clusters born within a tail environment that typically disperses before log age(yr) $\sim$ 8.5--9 \citep{binney08}, this is a suitable result. Shifting the model on the x-axis by 1 mag corresponds to a factor of 2.5 increase in cluster mass, so our color-magnitude criteria is sensitive to most appropriately-aged clusters $\gtrsim$ 5 $\times$10$^5$ \msun. Stochastic effects \citep{maiz09} and cluster disruption \citep{anders} will necessarily introduce some uncertainty.

This also submits an age (or more accurately, star formation history) bias into the analysis -- for tails with old ($\gtrsim$ 300 Myr) tidal debris, if they did not experience sufficiently recent episodes of cluster formation, clusters with masses $\lesssim$ 5 $\times$ 10$^4$ \msun\ would have faded below our detection threshold. Given the apparent preference for star formation in tidal debris to occur soon after their host galaxies' initial encounter (\citealp{chien10}; \citealp{dim08}, \citealp{m&h94}), cluster detection is favored in young (in our case, $\lesssim$ 250 Myr old) tails. These young tails have a larger mass range of detectable clusters than old tails for populations born at early dynamical ages. This would be less of an issue if all tails produced ample quantities of high-mass clusters. But depending on the debris, more massive clusters may be buried in the statistics of background subtraction, or may simply be absent if the cluster formation events of a tail were not powerful enough to populate the $\gtrsim$ 5 $\times$ 10$^5$ \msun\ end of the observed cluster mass function.  

The middle panel of \mbox{Figure B1} considers the observational implications, displaying the M$_{V606}$ distributions of sources for three tails. These histograms have been background-subtracted; i.e.\ the number of background sources was subtracted from the number of in-tail sources for each bin. The background-subtracted numbers of cluster candidates $N_{\rm{SCC}}$ are also presented. We plot model LFs suggested by \S 3.3 for these tails -- a power-law with a slope of -2 (purple lines), and -2.5 (yellow lines). Note that this choice of LF is motivated by statistical diagnostics (\mbox{Figure 5}); we do not infer the LF from the individual distributions, but apply these LFs to them, normalized to the total number of SCCs of each. V$_{606}$ distributions are subject to multiple generations of cluster formation and fading and strong stochastic effects, but the fits are qualitatively acceptable.  

AM 1054-325 is the most SCC-rich of the sample; it hosts many strong cluster candidates within the tidal tail boundary. Its cluster formation events were sufficiently powerful to populate the high mass end of the cluster mass function, and therefore the existing LF with sources well to M$_V \lesssim$ -12. NGC 2782E is more modestly populated, with dozens of sources detected around the M$_V$ = -8.5 line. Thus, a sizable fraction of these will be selected as SCCs with the current \Vhst\ and \VI\ criteria. Lastly, despite its complement of sufficiently blue sources (evident in the online \mbox{Figure 3.1}), NGC 1487E contains no sources bright enough to be classified as SCCs. If most of the objects in this tail are indeed star clusters, our statistical diagnostics insinuate that the cluster formation events that produced them were insufficiently powerful to generate clusters bright enough for SCC designation. In this case, this tail does in fact contain some clusters, but a cursory glance at sources brighter than conventional magnitude limits would miss them.    

Unfortunately, extending magnitude limits to detect fainter, older and/or less massive clusters is only plausible for the closest interacting galaxies. The in-tail completeness curves for the tails examined in the middle panel of \mbox{Figure B1} are provided in the final panel, translated to absolute magnitudes. The luminosity distances to the host galaxies are provided for reference. The turnovers or breaks in the faint end of the distributions of the middle panel correspond to the distance-defined completeness limits (variations in exposure times in \mbox{Table 1} provide ancillary effects). Were the tails not completeness-limited, the \Vhst\ distributions would continue to climb with fainter magnitudes and more SCCs would be uncovered.

Thus, only in the closest tails (NGC 1487E/W through NGC 2782E/W) would low-mass, $\sim$ 10$^4$ M$_{\odot}$, clusters be seen in our data. More distant tails (ESO 376- G 028 through NGC 1614N/S) would only allow detection of $\sim$ 10$^4$ M$_{\odot}$ clusters younger than $\sim$ 100 Myr, or to more massive ($\sim$ 10$^5$ M$_{\odot}$), luminous clusters at a broader age range (that is, unless the images had longer exposure times). Consequently, debris like NGC 4747 is fortunate in that it is close enough that in-tail cluster candidates may be detected at faint magnitudes (M$_{V,50\%}$ = -6.5). Such a tail with a ``weak" cluster formation history may appear devoid of clusters within the limits of our \textit{HST} photometry if it were placed at larger distances. In summary, tidal tail star cluster detection is heavily dependent on cluster formation histories/strengths, and distances of the tails themselves. Given these fundamental age and interaction constraints, not all tails can be expected to contain cluster populations to M$_V \lesssim$ -8.5.

\subsection{Controlling for Contamination by Field Stars}

It can be shown that single stellar contaminants pose few problems for some tails at the M$_V <$ -8.5 level. This magnitude limit may also be relaxed to M$_V$ = -6.5 for certain tails, provided they have few bright (M$_V <$ -9) sources and are reasonably local ($\lesssim$ 25 Mpc) for our image depths. NGC 1487W and NGC 520 are close enough to consider cluster candidates as faint as M$_V$ = -6.5, and do not have clusters at M$_V <$ -8.5, so we concentrate further discussion on these debris regions. All other proximate tails either have nonzero \SCC\ values at  M$_V <$ -8.5, or have \SCC\ = 0 within 2.5$\sigma$ for all magnitude ranges.

 It can be safely assumed that the contribution of main sequence stars in the range M$_V$ = -6.5 -- -8.5 is negligible; these early O-type, $\gtrsim$ 50 \msun\ stars would be extremely rare for tidal tails whose \Vhst-band luminosities are typically 3--4 orders of magnitude below that of their host galaxies. The only source of contamination on tail timescales $\lesssim$ 1 Gyr would therefore be red supergiant stars (RSGs) bluer than our adopted \VI\ criterion. To estimate the contribution of these, we first used Padova\footnotemark[6] isochrones to identify the minimum main sequence turnoff mass of stars that would evolve into sufficiently bright and blue RSGs. Translating this into the corresponding isochrone age $t_{\rm{iso}}$, we estimated the appropriate specific evolutionary fluxes $B_{\rm{spec}}(t_{\rm{iso}})$ of single stellar population models of \citet{BC03}; see \S 2.4. These values are on order of 10$^{-12}$ L$_{\odot}^{-1}$ yr$^{-1 }$ for M$_V$ = -7.5 -- -6.5 and $<$ 4 $\times$10$^{-13}$ L$_{\odot}^{-1}$ yr$^{-1 }$ for M$_V$ = -8.5.

\footnotetext[6]{http://stev.oapd.inaf.it/cgi-bin/cmd} 

The number of RSGs $N_{\rm{RSG}}$ that evolve to a certain \VI\ in a time t$_{\rm{evol}}$, produced by a population of luminosity $L$ can then be written as $N_{\rm{RSG}}  \approx B_{\rm{spec}}(t_{\rm{iso}}) \times L \times t_{\rm{evol}}$. Thus, to obtain one RSG, L $\approx (B_{\rm{spec}}(t_{\rm{iso}}) \times t_{\rm{evol}})^{-1}$. We estimate t$_{\rm{evol}}$ as $\sim$ 10$^6$ yr from the isochrones for a range metallicities 0.4 Z$_{\odot}$ to 2.5 Z$_{\odot}$. Combining these, we infer log($L/L_{\odot}$) $>$ 4.5--6.5 for the cutoff M$_V$ values -6.5 to -8.5.

We then find the absolute $V$-band magnitude for that $L$. Translating $L$ to M$_V$ requires applying the age-dependent bolometric correction for the population, which is uncertain with our limited data. By implementing the safest correction given the tail age constraint (i.e.\ one the would result in the faintest M$_V$), we find that SSPs with a total  M$_V \lesssim$ -9 are required to produce one RSG for an M$_V <$ -6.5 selection in tidal tails. For our M$_V <$ -7.5 and -8.5 limits displayed in \mbox{Figure 4}, the populations must be brighter than M$_V$ $\approx$ -10 and -15, respectively. Thus, if we use a magnitude threshold of -8.5 in culling sources, objects brighter than M$_V \approx$ -15 may be clusters that are both massive enough to have detectably bright RSGs and have survived disruption. Knowing how many of these bright clusters survived to the present, we may extrapolate how many of these clusters have evaporated into the tidal tail field and how many observable RSGs were left behind to contaminate in-tail source lists. 

It is clear that, since we find no sources brighter than M$_V$ = -15 in any tidal debris, and this magnitude limit is a very conservative estimate, our initial color-magnitude criteria for identifying SCCs is secure against in-tail singe stellar contamination. The same cannot be stated for fainter magnitudes, as raising the bar to M$_V$ = -7.5 and -6.5 requires accounting for sources that certainly exist in many tails at M$_V \lesssim$ -10 and -9, respectively. 

For the faint M$_V$ = -6.5 cutoff, it can be seen in the online CMDs of Figures 3.2--3.4 (NGC 1487W -- NGC 520) that there are $\leq$ 6 of these M$_V$  sources in the tails, depending on the debris. If $\sim$ 10\% of the original M$_V <$ -9 clusters have survived to the present day (following early cluster disruption histories of, e.g.\ \citealp{whitmore07}; \citealp{bastian08} and references therein), that means each tail is contaminated by by a number of RSGs $\sim$ 10 times the number of these bright SCCs. We do not consider secular evolution (e.g.\ two-body relaxation) in this simple estimate.

This introduces $\leq$ 60 contaminants for the M$_V <$ -6.5 source lists of NGC 1487W and NGC 4747, and 0 for NGC 520. Repeating the exercise for M$_V <$ -7.5 results in essentially the same number, 20--60 contaminants for the former two tails and 0 for the latter.  The number of predicted RSGs in NGC 1487W and NGC 4747 is on par with the number of sources detected, and when compounded with the present counting statistics, precludes successful determination of clusters at M$_V <$ -6.5 to -7.5. Because the TDG of NGC 520 has no sources brighter than M$_V$ = -9, we can be certain of the \SCC\  excesses seen in \mbox{Figure 4} for the fainter detection limits. Unfortunately, This WFPC2 pointing reveals clustered star formation in the TDG and not in the NGC 520 tail itself, so nothing can be definitively said of the debris.

 It must be noted, however, that this does not mean there are $no$ clusters with M$_V \approx$ -6.5 -- -8.5 in tails close enough for these objects to be detected, but that their presence statistically cannot be confirmed with only $V$- and $I$-band data. We choose not to extend M$_V$ further to avoid higher risk of contamination and completeness effects.



\end{document}